\documentclass[journal, 11pt, onecolumn, draftclsnofoot]{IEEEtran}

\usepackage{bm}
\usepackage{amsthm}
\usepackage{amsmath}
\usepackage{amssymb}
\usepackage{graphicx}
\usepackage[left=1.25in, right=1.25in, top=1in, bottom=1in]{geometry}

\usepackage{hyperref}
\usepackage{color}
\usepackage{authblk}
\usepackage{caption}
\usepackage{subcaption}

\newcommand{\revise}[1]{{\color{black}{#1}}}

\newcommand{\R}{\mathbb{R}}

\newcommand{\N}{\mathcal{N}}
\newcommand{\E}[1]{\mathbb{E}\left\{{#1}\right\}}

\numberwithin{equation}{section}
\numberwithin{thm}{section} % important bit

\begin{document}

\title{Single-particle cryo-electron microscopy: Mathematical theory, computational challenges, and opportunities}
\author{Tamir Bendory, Alberto Bartesaghi, and Amit Singer}
\date{}

\maketitle

\begin{abstract}
	
In recent years, an abundance of new molecular structures have been elucidated  using cryo-electron microscopy (cryo-EM), largely due to  advances in hardware technology and data processing techniques. 
Owing to these new exciting developments, cryo-EM was selected by Nature Methods as Method of the Year 2015, and the Nobel Prize in Chemistry 2017 was awarded to three pioneers in the field. 

The main goal of this article is to introduce the challenging and exciting computational tasks involved in reconstructing 3-D molecular structures by cryo-EM.
Determining molecular structures requires a wide range of computational tools in a variety of fields, including signal processing, estimation and detection theory, high-dimensional statistics, convex and non-convex optimization, spectral algorithms, dimensionality reduction,  and machine learning.
The tools from these fields must be adapted to work under  \revise{exceptionally challenging} conditions, including extreme  noise levels, the presence of missing data, and massively large datasets  \revise{as large as 
} several Terabytes.

In addition, we present two statistical models: multi-reference alignment and multi-target detection, that abstract away much of the intricacies of cryo-EM, while retaining some of its essential features. 
Based on these abstractions, we discuss some recent intriguing results in the mathematical theory of cryo-EM, and delineate relations with group theory, invariant theory, and information theory.
%We conclude the article by listing the computational and theoretical challenges that still lie ahead.

\end{abstract}

\newpage 

%-------Introduction--------------------

\section{Introduction}

Structural biology studies the structure and dynamics of macromolecules in order to broaden our knowledge about the mechanisms of life and impact the drug discovery process.
Owing to recent ground-breaking developments, chiefly in hardware technologies and  data processing techniques, an abundance of new molecular structures have been elucidated to near-atomic resolutions using cryo-electron microscopy (cryo-EM)~\cite{kuhlbrandt2014resolution,liao2013structure,bartesaghi20152,bai2015cryo,vinothkumar2016single,nogales2015cryo}. 
Cryo-EM was selected by Nature Methods as Method of the Year 2015, and the Nobel Prize in Chemistry 2017
was awarded to Jacques Dubochet, Joachim Frank and Richard Henderson ``for developing cryoelectron
microscopy for the high-resolution structure determination of biomolecules in solution\footnote{\revise{\url{https://www.nobelprize.org/prizes/chemistry/2017/press-release/}}}.''

\begin{figure}
	\centering 
	\includegraphics[width=1\linewidth]{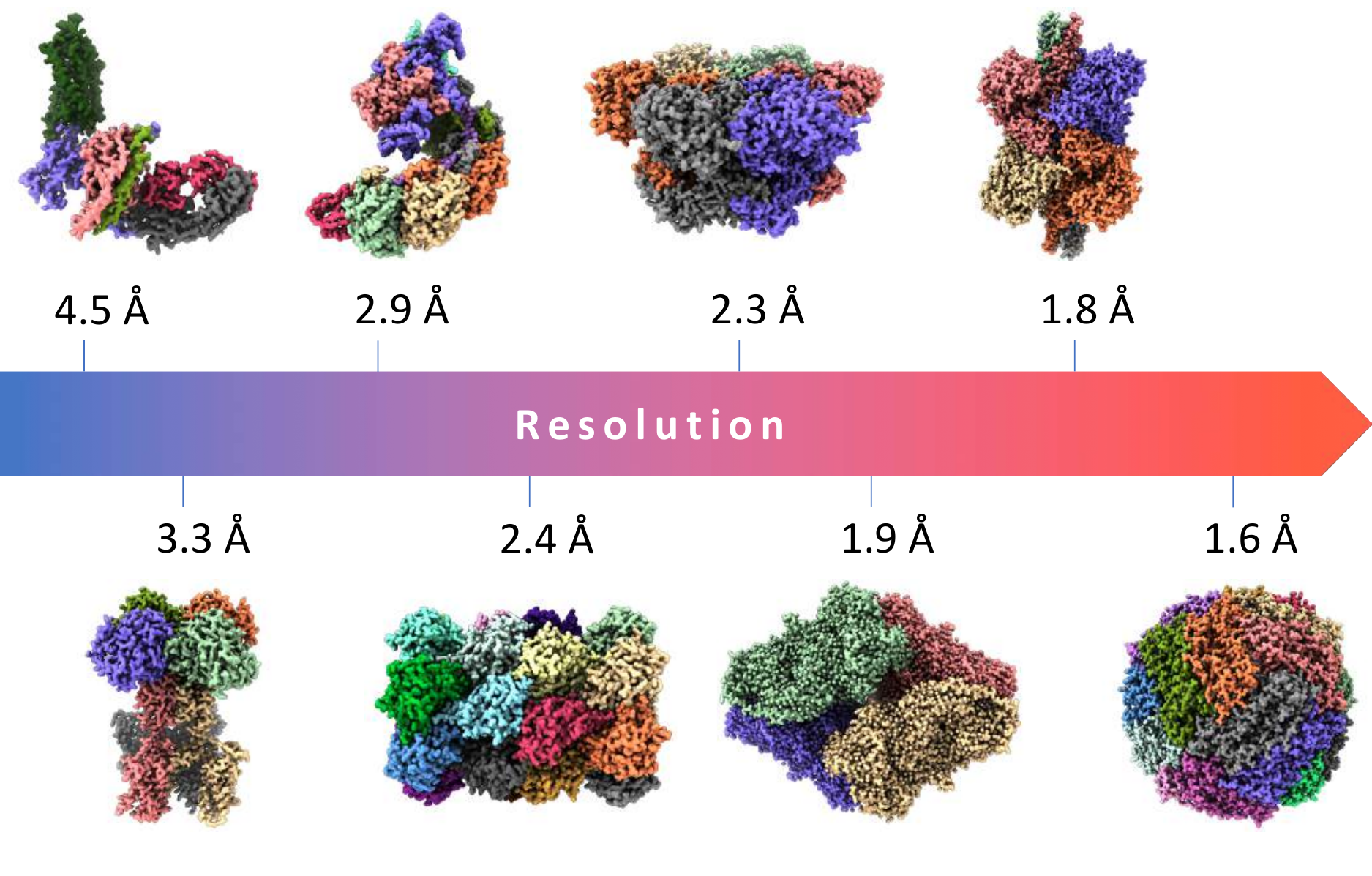}
	\caption{Gallery of important biomedical structures solved by single-particle cryo-EM at increasing resolutions. From left to right (lower to higher resolution): 4.5\AA~structure of the human rhodopsin receptor bound to an inhibitory G protein~\cite{kang2018cryo}, a member of the family of G-protein coupled receptors which are the target of $\sim 35\%$ of FDA-approved drugs; 3.3\AA~map of a voltage-activated potassium channel, an integral membrane protein responsible for potassium ion transport~\cite{matthies2018single}; 2.9\AA~cryo-EM structure of a CRISPR-Cas Surveillance Complex implicated in gene editing~\cite{guo2017cryo}; 2.4\AA~map of the T20s proteasome, a complex that degrades unnecessary or damaged proteins by proteolysis (data available from EMPIAR-10025)~\cite{empiar}; 2.3\AA~structure of human p97 AAA-ATPase, a key mediator of several protein homeostasis processes and a target for cancer~\cite{banerjee20162}; 1.9\AA~structure of $\beta$-galactosidase enzyme in complex with a cell-permeant inhibitor (EMPIAR-10061)~\cite{bartesaghi2018atomic}; 1.8\AA~structure of the conformationally dynamic enzyme glutamate dehydrogenase~\cite{merk2016breaking}; and 1.6\AA~map of human apoferritin, a critical intracellular iron-storage protein (EMPIAR-10200).}
	\label{fig:gallery}
\end{figure}

\begin{figure}
	\centering 
	\includegraphics[width=1\linewidth]{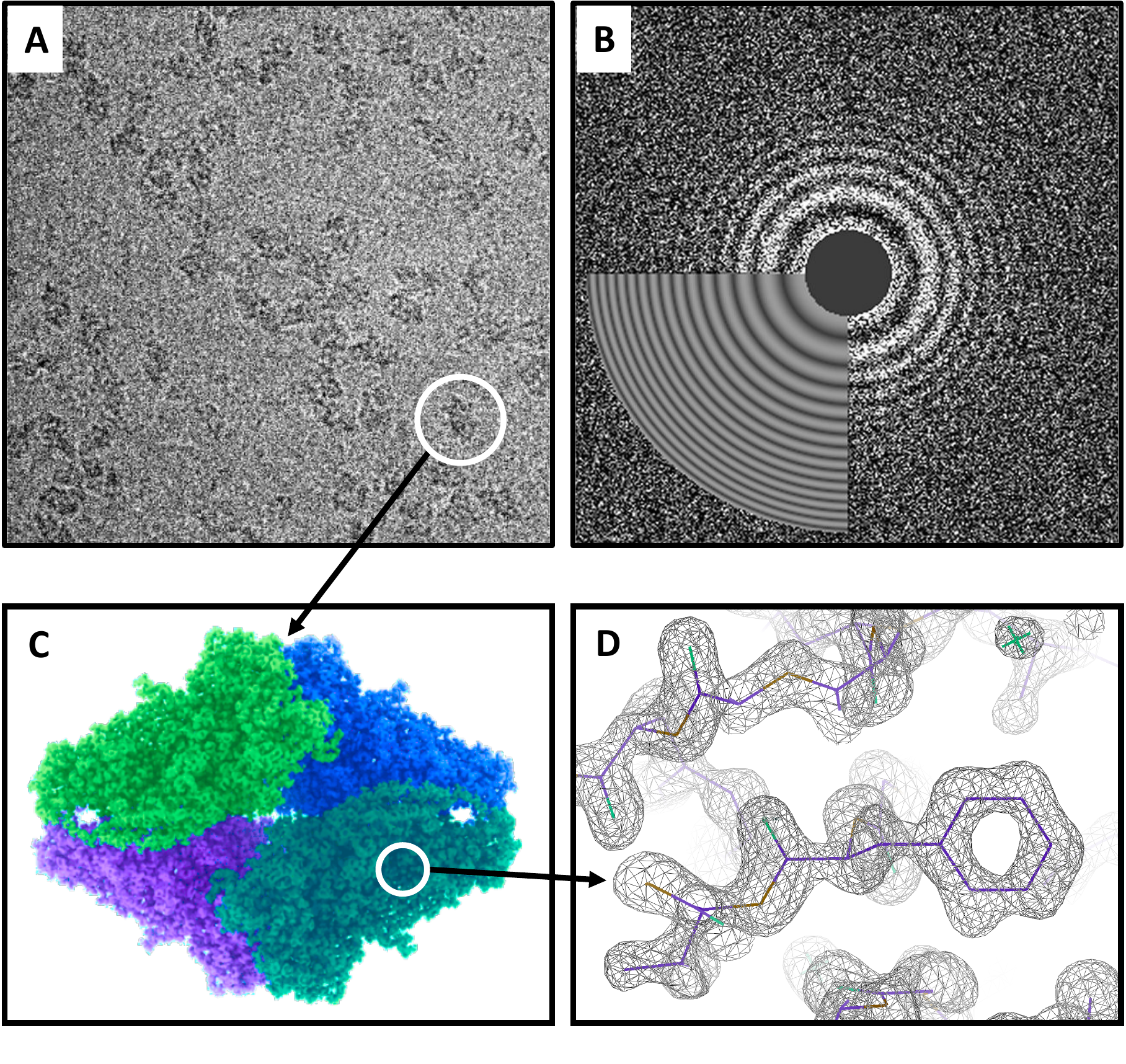}
	\caption{ High-resolution cryo-EM imaging of the $\beta$-galactosidase enzyme in complex with a cell-permeant inhibitor. A) Micrograph of $\beta$-galactosidase showing individual particle projections (indicated with white circle). B) Power spectra of image shown in \revise{panel A} and estimated contrast transfer function (CTF) matching the characteristic Thon ring oscillations (see Section~\ref{sec:CTF}). C) 1.9\AA~resolution map obtained from $\sim150,000$ individual particle projections extracted from the publicly available dataset EMPIAR-10061~\cite{bartesaghi20152,bartesaghi2018atomic}. D) Close-up view of reconstruction shown in \revise{panel C} highlighting high-resolution features of the map at the individual amino acid level.}
	\label{fig:1}
\end{figure}

In a cryo-EM experiment, \revise{biological macromolecules suspended in a liquid solution are} rapidly frozen into a thin ice layer.  The 3-D location and orientation of particles within the ice are random and unknown. 
An electron beam then passes through the sample and a 2-D tomographic projection, called micrograph, is recorded. 
The goal is to reconstruct \revise{a high-resolution estimate of the } 3-D electrostatic potential of the molecule (\revise{in particular, its atomic structure}) from a set of micrographs.
\revise{The {resolution} measures the smallest detail that is distinguishable in a recovered  3-D structure; structures with better resolutions resolve finer features. For example, at resolutions of $9\text{\AA}$ $\alpha$-helices are resolved;
at resolutions  of $4.8 \text{\AA}$ individual $\beta$-strands are resolved; 
at resolutions of  $3.5 \text{\AA}$ many amino acids side-chains are resolved~\cite{bai2015cryo}.}
\revise{Figure~\ref{fig:gallery} shows a gallery of important biomedical structures solved by cryo-EM at increasingly higher resolutions.}
Figure~\ref{fig:1} presents an example of a micrograph \revise{of the enzyme $\beta$-Galactosidase} and the corresponding high-resolution 3-D reconstruction~\cite{bartesaghi2018atomic}.

The signal-to-noise ratio (SNR) of cryo-EM data is very low due to two compounding reasons. On the one hand, the micrograph's contrast is low because of the absence of contrast enhancement agents, such as heavy-metal stains. On the other hand,  the noise level is high because the electron doses must be kept low to prevent damage to the radiation sensitive biological molecules. 
The difficulty to estimate the 3-D structure in this low SNR regime, when the  orientation and location of the particles are unknown, is the crux of the cryo-EM problem.

\revise{Forty years ago,}  Dubochet and colleagues devised a new technique to preserve biological samples within a thin layer of \revise{an amorphous solid form of water, called \emph{vitreous ice}~\cite{dubochet1982electron}.
In contrast to ``regular ice'', vitreous ice  lacks crystalline  molecular arrangement and therefore allows preservation of biological structures.}
As the  vitreous ice is maintained  at \revise{liquid nitrogen or liquid helium}  temperatures, the technique  was named ``cryo-EM''.
In the next decades, the successful application of cryo-EM  was limited to study  large and highly symmetric structures, such as ribosomes and different types of viruses. 
Before 2013, only few structures were resolved at resolutions better than  7\AA~and the field was dubbed ``blob-ology'' due to the blobby appearance of structures at these resolutions. 
We refer the reader to~\cite{bai2015cryo,nogales2015cryo,vinothkumar2016single,nogales2015development,frank2017advances} for a more detailed historical account  \revise{of the development of the technology.}

\begin{figure}
	\centering 
	\includegraphics[width=1\linewidth]{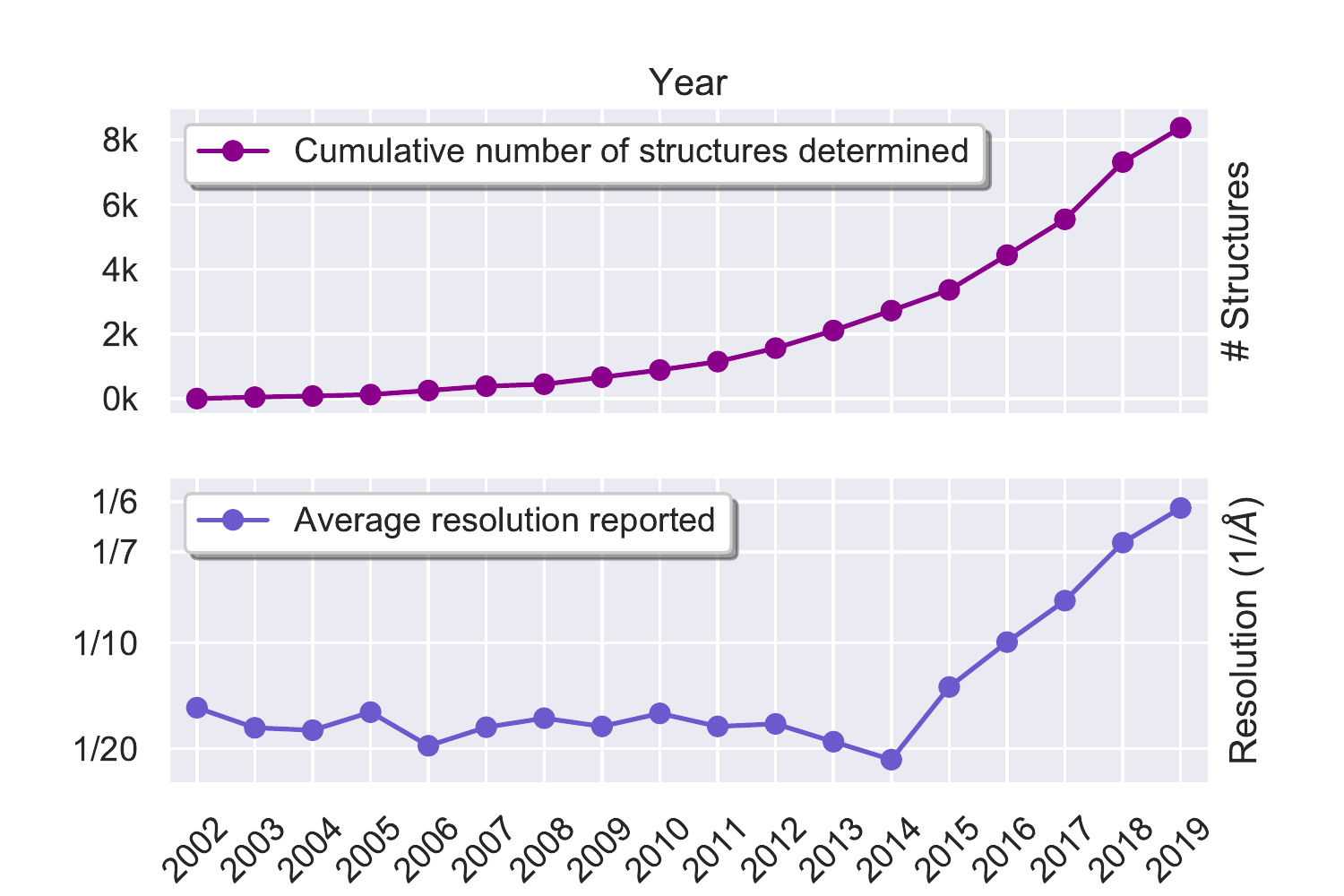}
	\caption{Recent growth in the number of high-resolution structures produced by single-particle cryo-EM. A) Cumulative number of structures solved by single-particle cryo-EM in the last 17 years as recorded in the Electron Microscopy Data Bank (EMDB) public repository~\cite{emdb}. B) Corresponding values for the average resolution of maps deposited in the database showing an inflection point after the year 2013, coinciding with the introduction of direct electron detectors technology in the cryo-EM field.}
	\label{fig:structures_determined}
\end{figure}

Since 2013, single particle reconstruction using cryo-EM \revise{has been undergoing fast} transformations, leading to an abundance of \revise{new} high-resolution structures, reaching close to atomic resolution~\cite{bartesaghi2018atomic}.
Figure~\ref{fig:structures_determined} presents the historical growth in the number of high-resolution structures produced by cryo-EM.
Several factors contributed to the ``resolution revolution''~\cite{kuhlbrandt2014resolution}. 
First and foremost, a new generation of detectors was developed, called \revise{direct electron detectors~(DED),  
that---in contrast to CCD cameras---do not convert electrons to photons. The DED}
dramatically 
improved  image quality \revise{and SNR}, and thus 
increased the attainable resolution of cryo-EM.
These detectors have high output frame rates that allow to record  multiple frames per micrograph (``movies'') rather than integrate individual exposures~\cite{ripstein2016processing}. 
\revise{These movies can be} used to compensate for \revise{motion induced  by the electron beam to the sample}; see Section~\ref{sec:motion_correction}.
In addition, recent hardware developments have enabled the acquisition and storage of huge amounts of data, which combined with readily access to CPUs and GPU resources, have helped propel the field forward.

The prevalent technique in structural biology \revise{in the last half a century} was X-ray crystallography. This technique suffers from three intrinsic weaknesses that can be mitigated by using cryo-EM imaging. First, many molecules, among them different types of membrane proteins, \revise{were notoriously difficult to crystallize.} 
In contrast, the sample preparation procedure for a cryo-EM experiment is significantly simpler, \revise{does not require crystallization,} and \revise{needs} smaller amounts of sample. 
Second, \revise{crystal contacts may alter the structure of proteins making it difficult to recover their physiologically relevant conformation. Cryo-EM samples, instead, are rapidly} frozen in vitreous ice which  preserves the molecules in a near-physiological state. 
Third, the X-ray beam aggregates the information from all molecules simultaneously and thus hinders visualization of \revise{structural} variability. In stark contrast, the projection of each particle in a cryo-EM experiment  is recorded independently and thus
multiple \revise{structures}---associated with different functional states---can be estimated~\cite{sorzano2019survey}.
\revise{A third technique,} nuclear magnetic resonance (NMR), can be used to elucidate molecular structures  in physiological conditions (water solution at room temperature) but is restricted to small structures, up to~$\sim$50~kDa.

The main goal of this article is to introduce the unique and exciting algorithmic challenges and cutting-edge mathematical problems arising from cryo-EM research. 
Estimating  molecular structures involves developing and adopting computational tools in signal processing, estimation and detection theory, high-dimensional statistics, convex and non-convex optimization, spectral algorithms, dimensionality reduction,  and machine learning, as well as  knowledge in   group theory, invariant theory, and information theory.
All tools  from the aforementioned fields should be adapted to  exceptional conditions: an \revise{extremely} low SNR environment and the presence of missing data (for instance, \revise{2-D} location and 3-D orientation of samples in the micrograph). 
In addition, the devised algorithms should be efficient in order to run on massively large  datasets  in the order of several Terabytes. 
The article  provides an account of the  leading software packages in the field and discusses their underlying mathematical, statistical, and algorithmic principles~\cite{scheres2012relion,tang2007eman2,punjani2017cryosparc,grant2018cistem}.

The article is organized as follows. In Section~\ref{sec:inverse_problem}, we describe the generative (forward) model of cryo-EM images and formulate the inverse problem of estimating the 3-D structure of \revise{molecules}.  Section~\ref{sec:computional_challenges} stresses the main computational challenges in cryo-EM data processing: high noise level, missing data, and massive datasets.  
Section~\ref{sec:3D_reconstruction} presents 3-D reconstruction methods. In particular, Section~\ref{sec:high-res} discusses high-resolution techniques and  elaborates upon the maximum likelihood framework, which has been shown to be a highly effective tool.  Then, Section~\ref{sec:ab_initio_modeling} introduces different methods for \emph{ab initio} modeling, frequently used to constitute 3-D low-resolution estimates. 
Section~\ref{sec:building_blocks} describes different building blocks in the algorithmic pipeline of cryo-EM, and relates them to detection, classification, denoising, and dimensionality reduction techniques. Section~\ref{sec:abstract_models} introduces recent results concerning two mathematical models: multi-reference alignment and multi-target detection. These models  abstract the cryo-EM problem  and provide a broader analysis framework. 
In Section~\ref{sec:challenges_ahead} we suggest future research directions and underscore important---computational and theoretical---open questions.
Section~\ref{sec:perspective} concludes the article.

Before moving on, we want to mention that topics which are of great importance to practitioners, such as the physics and optics of an electron microscope, sample preparation, and data acquisition \emph{are not discussed} in this article. 
These topics are thoroughly covered by biological-oriented surveys, such as~\cite{vinothkumar2016single,nogales2015cryo,nogales2015development,frank2017advances} and references therein.

%-------The cryo-EM inverse problem--------------------

\section{The cryo-EM \revise{reconstruction} problem}
\label{sec:inverse_problem}

\begin{figure}
	\centering 
	\includegraphics[width=1\linewidth]{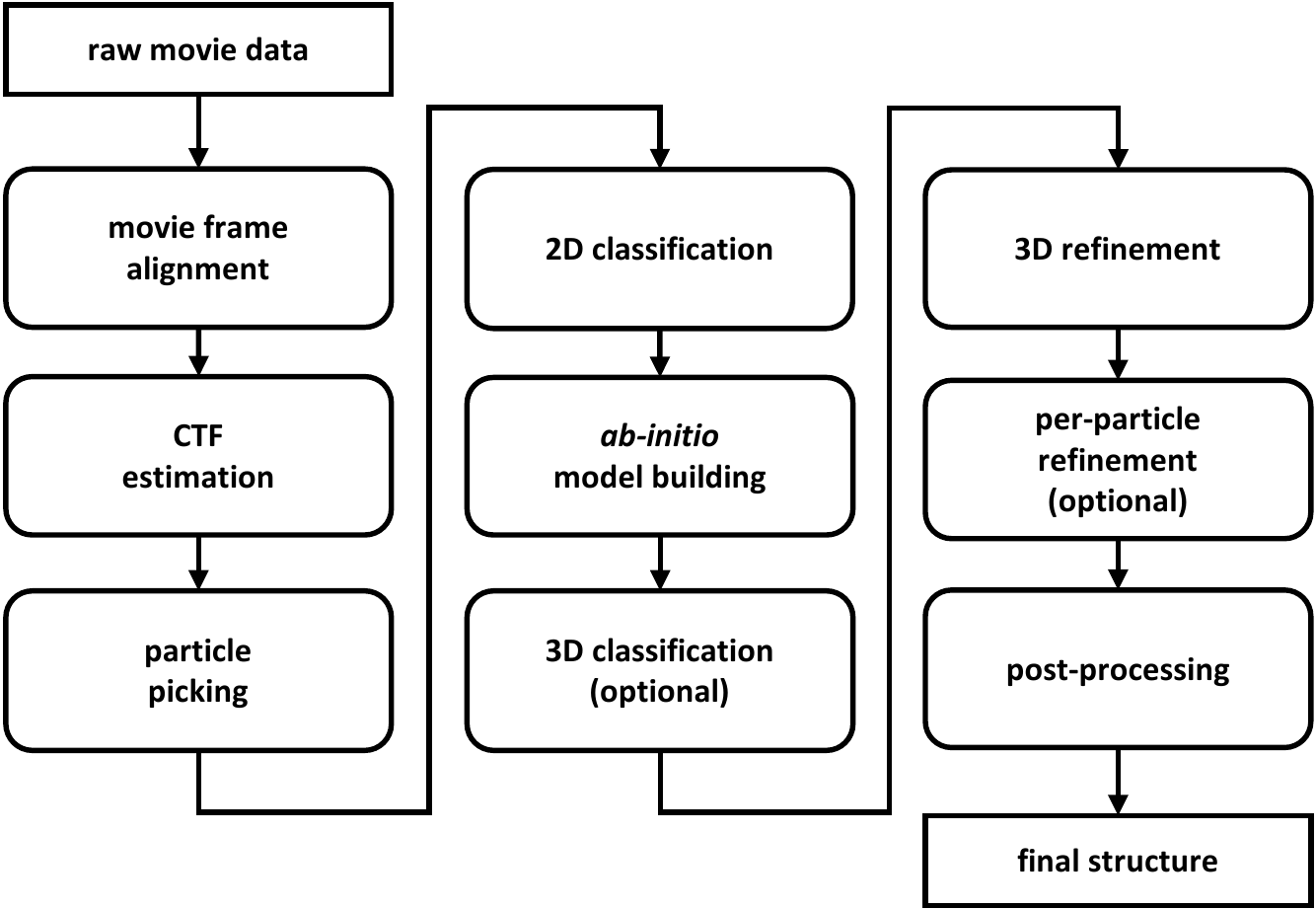}
	\caption{\revise{Flowchart diagram showing the computational pipeline required to convert raw movie data into high-resolution structures by single-particle cryo-EM. Raw data is first pre-processed at the movie frame alignment (Section~\ref{sec:motion_correction}) and CTF estimation (Section~\ref{sec:CTF}) steps followed by particle picking  to detect and extract the  individual projections from micrographs. Occasionally, particle picking is followed by a particle pruning stage to remove non-informative picked particles  (Section~\ref{sec:particle_picking}).
	The output of this stage is a set of 2-D images, each (ideally) contains a noisy tomographic projection taken from an unknown viewing direction.
	Particle images are then classified (Section~\ref{sec:2Dclassification}): the 2-D  classes are used to construct \emph{ab initio} models, as templates for particle picking, to provide a quick
	assessment of the particles, and for symmetry detection. Then, an initial, \emph{ab initio} model is built using the 2-D classes or by alternative techniques (Section~\ref{sec:2Dclassification}). 
	If the data contains  structural variability or a mix of structures (see Sections~\ref{sec:inverse_problem} and~\ref{sec:heterogeneity}), then  a 3-D classification step is applied   to cluster the projection images into the different structural conformations. Initial models are subjected to 3-D high-resolution refinement (Section~\ref{sec:high-res}) and an additional per-particle refinement may be applied. Finally, a post-processing stage is employed to facilitate interpretation of structures in terms of atomic models. Different software packages  may use slightly different workflows and occasionally some of the steps are applied iteratively. For instance, one can use the 2-D classes to repeat particle picking with more reliable templates. This figure is adapted from~\cite{zhou2019unsupervised}.}} 
	\label{fig:flowchart}
\end{figure}

\revise{Modern electron microscopes produce} multiple micrographs,  each one of them  composed of a series of frames (a ``movie''). The first stage of any contemporary algorithmic pipeline is to align and average the frames in order to mitigate the effects of \revise{movement induced by the electron beam}, and thus to improve the SNR. This process is called motion correction\revise{, or movie frame alignment}.  The next step, termed particle picking, consists in detecting and extracting the particles' tomographic projections from the micrographs. 
\revise{Perfect detection requires finding the particles' center of mass, which is difficult to estimate due to the characteristics of the Fourier transform of the microscope's point spread function (PSF)---called the contrast transfer function (CTF); see Section~\ref{sec:CTF}. } 
The output of \revise{the particle picking} stage is a series of images $I_1,\ldots,I_N$ from which we wish to estimate the 3-D structure; 
\revise{this section focuses on the latter problem of  3-D reconstruction from picked particles, which is the heart of the computational pipeline of single-particle reconstruction using cryo-EM.}
Motion correction and particle picking are discussed in more detail in Section~\ref{sec:building_blocks}.
\revise{Figure~\ref{fig:flowchart} shows the complete cryo-EM imaging pipeline.}

Let $\phi: \R ^3\rightarrow\R$ represent the  3-D  molecular structure to be estimated. 
Under the assumption that the particle picking is executed  well \revise{(namely, each particle is detected up to a small translation; we assume for simplicity no false detection)}, each image $I_1,\ldots,I_N$ is formed by rotating~$\phi$ by a 3-D rotation $R_\omega$, integrating along the z-axis (tomographic projection), shifting by a 2-D shift $T_t$, convolving with the PSF  of the microscope \revise{$h_i$},  and adding noise:
\begin{align} \label{eq:model_explicit}
I_i &= h_i\ast T_{t_i}\left(\int_{-\infty}^\infty\phi\left(R_{\omega_i}^Tr\right)dz\right) +  \text{``noise''},   &r= (x,y,z), \,\, i=1,\ldots,N,
\end{align} 
where $\ast$ denotes convolution.
The model can be written more concisely as
\begin{align} \label{eq:model}
I_i &= h_i\ast T_{t_i} P R_{\omega_i}\phi +  \text{``noise''},  & i=1,\ldots,N,
\end{align} 
where the  tomographic projection is denoted by $P$, and $(R\phi)(r) = \phi(R^Tr)$. 
The image is then sampled on a Cartesian grid. 
The integration along the z-axis is called the X-ray transform (not to be confused with the Radon transform in which the integration is over hyperplanes)~\cite{natterer1986mathematics}. 
The 3-D rotations describe the  unknown 3-D orientation of the particles embedded in the ice, and can be thought of as random elements of the special orthogonal group $\omega\in SO(3)$. These rotations can be represented as $3\times 3$ orthogonal matrices with determinant one or by quaternions. 
The 2-D shifts $t\in\R^2$ result from detection inaccuracies, which are usually small.  
The PSFs \revise{$h_i$} are assumed to be known and their Fourier transforms suffer from many zero-crossings, making  deconvolution  challenging; see Section~\ref{sec:CTF} for further discussion.

\begin{figure}
	\centering 
	\includegraphics[width=.8\linewidth]{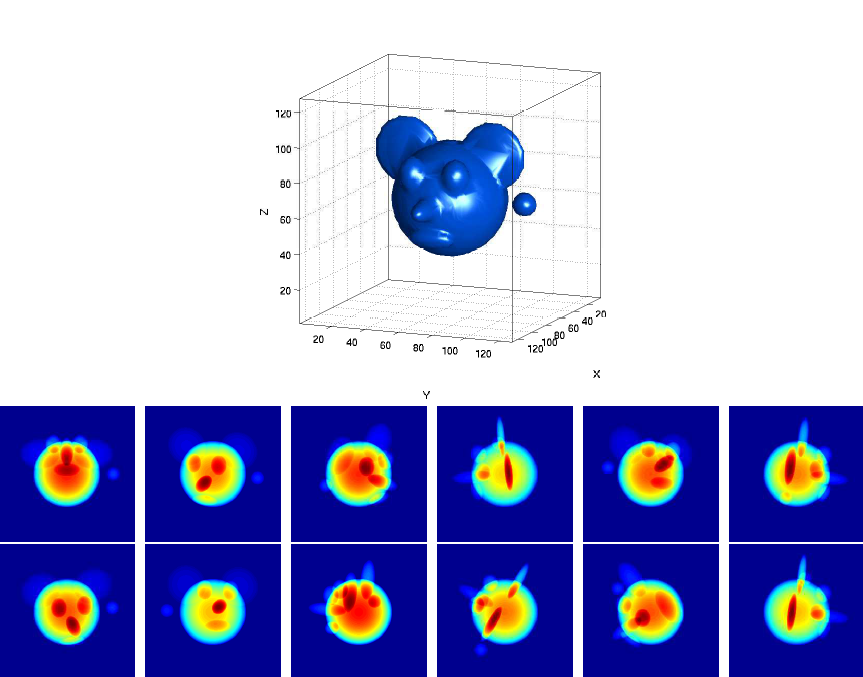}
	\caption{A simulated 3-D structure and a dozen of its noise-free tomographic projections from different viewing directions. The most simplified version of the cryo-EM problem is estimating the 3-D structure from the 2-D projection images when the viewing directions are unknown. In practice, the projections are highly noisy, slightly shifted, and convolved with the microscope's PSF. Image credit:~\cite{singer2018mathematics}.}
	\label{fig:clean_images}
\end{figure}

The cryo-EM inverse problem \revise{of reconstruction} consists in estimating the 3-D structure $\phi$ from the 2-D images $I_1,\ldots,I_N$.
Importantly, the 3-D rotations $\omega_1,\ldots,\omega_N$ and the 2-D shifts  $t_1,\ldots,t_N$ are \emph{nuisance variables.} Namely,  while the rotations and the shifts are unknown a priori, their estimation is not an aim by itself.
Figure~\ref{fig:clean_images} shows an example of
the cryo-EM problem in its most simplified version, without noise, CTF, and shifts.

The reconstruction of $\phi$ is possible up to \revise{three} intrinsic ambiguities: a global 3-D rotation, \revise{the position of the center of the molecule (3-D location)}, and  handedness.
\revise{The latter symmetry, also called chirality, means that it is impossible to distinguish whether the  molecule was reflected
	about a 2-D plane through the origin.
 The handedness of the structure} cannot be determined \revise{from} cryo-EM images alone because the original 3-D
object and its reflection give rise to identical sets of projections related by the following conjugation $\tilde{R}_{\omega_i} = JR_{\omega_i}J^{-1}$, where $J=\text{diag}(1,1,-1)$.

In the presence of structural variability, $\phi$ may be thought of as a random signal with an unknown distribution defined over a space of possible structures (which might be unknown as well). 
In this case, the task is more ambitious and involves estimating the whole distribution of conformations. 
Usually, the distribution is assumed to be discrete (that is, in each measurement we observe one of few possible structures \revise{or conformational states}) or to lie in a low-dimensional subspace or manifold. 
This subject is further discussed in Section~\ref{sec:heterogeneity} and in a recent survey~\cite{sorzano2019survey}.

The chief noise source in cryo-EM \revise{at the frame level (before motion correction)} is shot noise, which follows a Poisson distribution~\cite{sigworth2004classical}. 
\revise{After movie frames are averaged to produce micrographs, it is customary to assume that the noise is characterized by a  Gaussian distribution.}
Indeed, all current algorithms build upon---implicitly or explicitly---the speculated Gaussianity of the noise. While the spectrum of the noise is not white---mainly due to inelastic scattering, variations in the thickness
of ice, and the PSF of the microscope---it is assumed to be a 1-D radial function (namely, constant along the angular direction). The common practice is to estimate the parameters of the noise power spectrum \revise{during the 3-D reconstruction process, or}
  from  regions in the micrographs that, presumably, contain no signal. 
	
%-------Main computational challenges--------------------

\section{Main computational challenges}
\label{sec:computional_challenges}

The difficulty to determine high-resolution molecular structures using cryo-EM hinges upon three characteristic features of the cryo-EM data: high noise level, missing data, and massive datasets. This section elaborates on these unique features, while the next sections dive into the different tasks and algorithms involved in cryo-EM data processing.

\paragraph{High noise level} In a cryo-EM experiment, the electron doses must be kept low to  mitigate radiation damage due to the electron illumination. 
The low doses induce high noise levels on the \revise{acquired raw data frames}. Together with the image's low contrast, it results in SNR levels that are usually well below 0 dB and might be as low as \revise{-20 dB} (namely, the power of the noise is 100 times greater than the power of the signal). 
Under such low SNR conditions, standard tasks, such as aligning, detecting, or clustering signals, become very challenging. 

To comprehend the difficulty to perform estimation tasks in a low SNR environment, let us consider a simple rotation estimation problem.
Let us denote by $x\in\mathbb{R}^L$  the samples of a {periodic} signal  on the unit circle; $x$ is assumed to be known.
We wish to estimate a rotation $\theta\in[0,2\pi)$  from  a noisy measurement
\begin{align} \label{eq:alignment}
y  = R_{\theta} x + \varepsilon, 
\end{align}
where $R_{\theta}$ denotes the rotation operator and $\varepsilon\sim \mathcal{N}(0,\sigma^2I)$.
To estimate  $\theta$,  we correlate the signal $x$ with rotated versions of $y$, and choose the inverse of the rotation that maximizes the correlation as the estimator---this technique is called \emph{template matching}. 
In the absence of noise, template matching simply correlates  the signal with its rotated versions; the maximum is attained  when $y$ is rotated by $-\theta$. However, in the presence of noise, we get an additional stochastic term  due to the correlation of the noise with the signal. Consequently, if $\sigma^2$ is large, it is likely that the peak of the correlation will not be close to  $-\theta$. In particular, when $\sigma\to\infty$ the location of the peak is distributed uniformly on the circle.  
This result can be derived  formally using the Neyman-Pearson lemma and holds true for any estimation technique, not necessarily template matching~\cite{bendory2019multi}.  
Even if we collect $N$ measurements (each with a different rotation), it is impossible to estimate the $N$ rotations accurately. 
In~\cite{aguerrebere2016fundamental}, it was shown that the Cram\'er-Rao bound of this problem is proportional to $\sigma^2$ and independent of~$N$. Therefore, if the noise level is high, the variance of any estimator would be high as well. 

\revise{The same conclusions that were derived for the simple rotation estimation problem~\eqref{eq:alignment} remain true for cryo-EM. For example, even if the 3-D structure is known, aligning a noisy raw image against multiple noiseless projection templates sampled from $SO(3)$ will produce no salient peak in the correlation if the noise level is high. In particular, the higher the noise level, the flatter the distribution over~$SO(3)$.}

\paragraph{Missing data---Unknown viewing directions and locations} The viewing direction and location associated \revise{with each particle} in a micrograph are unknown a priori. If they were known, estimating the structure $\phi$ would be a linear inverse problem---similarly to the reconstruction problem in computerized-tomography (CT). The recovery in this case is based on the \emph{Fourier slice theorem}, stating that the 2-D Fourier transform of a tomographic projection is the restriction of the 3-D Fourier transform of $\phi$ to a 2-D plane. \revise{Mathematically, it can be written succinctly as
\begin{equation}
S R F_3 \phi = F_2 P R \phi,
\end{equation}
where $F_2$, $F_3$, $P$  and $S$ are, respectively, the 2-D Fourier, 3-D Fourier, tomographic projection and restriction} operators; this theorem is a direct corollary of the fact that \revise{the Fourier operator $F$  and the rotation operator $R$} commute, i.e., $RF=FR$.

The Fourier slice theorem implies that acquiring  tomographic projections from known viewing directions is equivalent to sampling the 3-D Fourier space. Hence, given enough projections, one can estimate $\phi$ to a certain resolution. This is the underlying principle behind many CT imaging algorithms, such as \revise{the classical} filtered back projection (FBP)~\cite{natterer1986mathematics}.
\revise{However, FBP is not optimal for cryo-EM (even if viewing directions are known) since not all  viewing directions are necessarily represented in the data, and the sampling in  Fourier domain is non-uniform; the coverage of viewing directions affects the quality of the solution~\cite{baldwin2019non}. 
\revise{An alternative  is the algebraic reconstruction technique (ART) which solves the linear system of equations by iterative projections; this algorithm is called  the \emph{Kaczmarz method} in numerical linear algebra. However, ART is rarely used in cryo-EM because it is slow: it does not exploit the  Fourier slice theorem and thus  cannot be accelerated using FFTs.}
Modern approaches were developed to exploit the Fourier slice theorem and to account for 
non-uniform sampling; popular algorithms are based on efficient gridding methods (that compute a uniformly sampled version of a function  from a nonuniformly sampled version by choosing proper weights)~\cite{penczek2004gridding} and using non-uniform FFT packages  to harness the structure of the linear system~\cite{dutt1993fast,greengard2004accelerating,barnett2018parallel}}.

While the viewing directions in cryo-EM are unknown, there is a rigorous technique to estimate them based on common-lines; see Section~\ref{sec:common_lines} and~\cite{van1987angular,singer2011three}. 
Unfortunately, any method to estimate the viewing directions is destined to fail when the SNR is low, for the same reasons that estimating $\theta$ in~\eqref{eq:alignment} would fail.
Bearing in mind that the ultimate goal is to estimate the 3-D structure---not the viewing directions---it is essential to consider statistical methods that circumvent rotation estimation, such as maximum likelihood and the {method of moments}. 

\paragraph{Massive datasets \revise{and high dimensionality}} 

A single session of data collection in a typical cryo-EM experiment produces a few thousand micrographs, each containing several hundred individual particle projections. Depending on the type of detector used during acquisition, micrographs can range in size from a few tens of Megapixels up to one hundred Megapixels for the newest cameras. Moreover, the new generation of detectors can record each micrograph as a rapid burst of frames, producing large movie files that result in datasets of several Terabytes in size. The sheer volume of data must be taken into account early on in the algorithmic design process, and in addition to  storage considerations, steps must be taken to ensure the efficient use of computational resources in order to keep up with the ever-increasing throughput of data produced by modern cameras. 

\revise{Another issue is the dimensionality of the reconstruction problem. The number of voxels of a typical $200\times 200\times 200$ volume is 8 million. Estimating so many parameters poses a challenge---both from the computational complexity side (how to efficiently optimize; for example, how to find the maximum of the likelihood function of $8$ million variables?) and also on the statistical estimation front (see Section~\ref{sec:PCA}). The problem is even more severe when multiple structures, or even a continuum of structures, need to be estimated (see Section~\ref{sec:heterogeneity}).
Getting to high resolution is a major bottleneck  because of a compounding computational burden effect: as more parameters need to be estimated,  more data is required for their estimation, and the computation becomes severely more expensive and challenging.}

%-------3-D reconstruction--------------------

\section{3-D reconstruction from projections}
\label{sec:3D_reconstruction}

\subsection{High-resolution refinement}
\label{sec:high-res}

The reconstruction procedure of high-resolution 3-D  structures is usually split into two stages: constructing an initial low-resolution model, which is later refined by  applying an iterative algorithm (``refinement algorithm''); \revise{See Figure~\ref{fig:flowchart}.} This section is devoted to  refinements techniques, while different approaches to constitute low-resolution estimates using \emph{ab initio} modeling---initialization-free models---are discussed in the next section. 

Refinement techniques for cryo-EM can be broadly classified into two categories: hard and soft angular assignment methods.   
The hard assignment approach is based on template matching.
At each iteration, multiple projections are generated from the current estimate of the structure; the projections should, ideally,  densely cover $SO(3)$. 
Then, a single viewing direction is assigned to each experimental image based on the projection with which it correlates best. 
The angular assignment can be performed either in real or Fourier space. The advantage of working in  Fourier space is that it is not necessary to rotate the molecule:  projections	can be computed fast using off-the-shelf non-uniform FFT packages (e.g.,~\cite{dutt1993fast,greengard2004accelerating,barnett2018parallel}) as implied by the Fourier slice theorem. Moreover, using this representation, the CTF is simply a diagonal operator.
Once the viewing directions of all experimental images \revise{are} assigned, the 3-D structure is constructed using standard linear \revise{inversion} techniques; see discussion in Section~\ref{sec:computional_challenges}.  
The algorithm iterates between hard angular assignment and structure construction until convergence.  
While the quality of hard angular assignments may be influenced by the high noise levels, there are several examples of packages that follow this type of approach, e.g., EMAN2~\cite{tang2007eman2} and cisTEM~\cite{grant2018cistem}, among others.

A second class of strategies  are the soft assignment methods.
Similarly to hard assignment methods, in each iteration the experimental images are correlated with multiple \revise{template} projections. 
However, instead of choosing the best match,
a similarity score is given to each pair of experimental image and projection. These scores, also called weights or responsibilities, are computed according to the generative statistical model of the experimental images. 

The soft assignment methods are instances or variants of the well-known  expectation-maximization \revise{(ML-EM)}  algorithm~\cite{dempster1977maximum}. 
Notably, this relation classifies the 3-D reconstruction problem as a problem of maximum likelihood estimation or, more generally, a problem in Bayesian statistics, and thus provides  a solid theoretical and algorithmic framework. In particular, it enables to incorporate priors that essentially act as regularizers in the reconstruction process.
In the context of cryo-EM, \revise{ML-EM}  was first applied to 2-D cryo-EM images by Fred Sigworth~\cite{sigworth1998maximum}, and later  implemented for 3-D reconstruction by the software packages RELION~\cite{scheres2012relion} and cryoSPARC~\cite{punjani2017cryosparc}. 

In what follows, we describe the \revise{ML-EM} algorithm for a \revise{more general statistical} model, and then particularize it to the special case of cryo-EM.
Suppose we collect $N$ observations from the model:
\begin{align} \label{eq:EM_model}
y_i &= L_{\theta_i} x + \varepsilon_i, &i=1,\ldots,N,
\end{align}  
where $L_\theta$ is a linear transformation acting on the signal $x$, parameterized by a random variable $\theta$, and $\varepsilon\sim\N(0,\sigma^2I)$.
The goal is to estimate $x$, while  $\theta_1,\ldots,\theta_N$ are nuisance variables.
A typical assumption is that $x$ can be represented by a  finite number of coefficients \revise{(for instance, its Fourier expansion is finite)} and these coefficients were drawn from a Gaussian distribution with  mean $\mu$ and covariance matrix $\Sigma$.
The key for reliable estimation is to marginalize over the nuisance variables $\theta_1,\ldots,\theta_N$. Without marginalization (that is, when the goal is to jointly estimate $x$ and $\theta_1,\ldots,\theta_N$), the number of parameters grows with the number of measurements indefinitely, and thus the maximum likelihood estimator may be inconsistent. This phenomenon is exemplified by the Neyman-Scott ``paradox''~\cite{neyman1948consistent}. 

In cryo-EM, the transformation $L_\theta$  is the  operator described in~\eqref{eq:model}. This operator rotates the volume, computes its 2-D tomographic projection, applies a  2-D translation and convolution with the PSF; $\theta$ is drawn from a distribution defined over the 5-D space of  3-D rotations and 2-D translations. The distribution of $\theta$ is generally unknown and should be estimated as part of the \revise{ML-EM} algorithm. 

Let us denote $\mathbf{y}=(y_1,\ldots,y_N)$. The posterior distribution $p(x|\mathbf{y})$ of~\eqref{eq:EM_model} is proportional to the product of the prior $p(x)$ with the likelihood function
\begin{align} \label{eq:likelihood}
\mathcal{L}(x; \mathbf{y})=p(\mathbf{y}|x)&=\prod_{i=1}^N \sum_{\theta_\ell\in \Theta}p(y_i|x,\theta_\ell)p(\theta_\ell)\\ &= \frac{1}{(2\pi\sigma^2)^{M/2}}  \prod_{i=1}^N\sum_{\theta_\ell\in \Theta}p(\theta_\ell)e^{-\frac{1}{2\sigma^2}\|y_i-L_{\theta_\ell}x\|^2}, \nonumber
\end{align}
where $\Theta$ is a discrete space on which $\theta$ is defined, and $M$ is the length of each observation. 
If the entire posterior distribution could be computed, then one could obtain far more statistical information about $x$ than just maximizing the likelihood. For instance, the best estimator in the minimum mean square error (MMSE) sense is obtained by marginilzing over the posterior:
\begin{align}
\hat{x}_{\text{MMSE}}(\mathbf{y}) = \int_x  p(x|\mathbf{y})xdx = \mathbb{E}\{x|\mathbf{y}\}.  
\end{align}
Unfortunately, the entire posterior can rarely be computed, especially in big data problems, such as cryo-EM.
The \revise{ML-EM} framework provides a simple, yet frequently very effective, iterative method that tries to compute the maximum a posteriori  estimator (MAP), namely,  the maximal value of~$p(x|\mathbf{y})$.   

Each iteration of the \revise{ML-EM}  algorithm  consists of two steps. The first step (E-step) computes the expected value of the log of the posterior with respect to the nuisance variables, conditioned on the current estimate of $x$, denoted here by $x_t$, and the data:
\begin{align} \label{eq:Q}
Q(x|x_t) &= \mathbb{E}_{\theta|\mathbf{y},x_t}\left\{\log p(x|\mathbf{y},\theta)\right\} \\ 
& = \text{constant} -\frac{1}{2\sigma^2}\sum_{i=1}^{N}\sum_{\theta_\ell\in\Theta}w_{i,\ell}\|y_i - L_{\theta_\ell}x\|^2 + \log p(x), \nonumber
\end{align} 
where 
\begin{equation} \label{eq:em_weights}
w_{i,\ell} = p(\theta_i = \theta_\ell|y_i,x_t) = \frac{e^{\frac{-1}{2\sigma^2}\|y_i-L_{\theta_\ell} x_t\|^2 }}{\sum_{{\theta_\ell}\in\Theta}e^{\frac{-1}{2\sigma^2}\|y_i-L_{\theta_\ell} x_t\|^2 }}.
\end{equation}
If $x\sim\mathcal{N}(0,\Sigma)$, then $\log p(x)=-\frac{1}{2}x^T\Sigma^{-1}x$.

The second step, called the M-step, updates $x$ by 
\begin{equation}
x_{t+1} = \arg\max_x Q(x|x_t),
\end{equation} 
which is usually performed by 
setting the gradient of $Q$ to zero. If $x$ is assumed to be Gaussian, then the M-step reduces to solving a linear system of equations with respect to $x$. The E and M steps are applied iteratively.  
It is well-known that $p(x_{t+1}|\mathbf{y})\geq p(x_t|\mathbf{y})$; however, since the landscape of $p(x|\mathbf{y})$ is usually non-convex, the iterations are not guaranteed to converge to the MAP estimator~\cite{dempster1977maximum}. 
Usually, the \revise{ML-EM} iterations halt when $\frac{p(x_{t+1}|\mathbf{y}) - p(x_t|\mathbf{y})}{p(x_t|\mathbf{y})}$ is  smaller than some tolerance (but other criteria can be employed as well). The posterior distribution at each iteration can be evaluated  according to~\eqref{eq:likelihood}. 

The implementation details of the \revise{ML-EM} algorithm for cryo-EM vary across different software packages. The popular package RELION, for example,  incorporates a prior of uniform distribution of rotations over $SO(3)$ (although the distribution itself is usually non-uniform),  and that each Fourier coefficient of the 3-D volume was drawn independently from a normal distribution~\cite{scheres2012relion}. 
%The translations are assumed to be distributed uniformly in a bounded range (specified by the user). 
%In addition, it incorporates a prior that each Fourier coefficient of the 3-D volume was drawn from an i.i.d.\ Gaussian variable. 
The variance of the \revise{Fourier coefficients'} prior is updated at each iteration by averaging over concentric frequency shells of the current estimate (so it is a 1-D radial function).
If the \revise{ML-EM} algorithm is initialized  with a smooth volume (which is the common practice), the variance would be large only at low-frequencies, hence high frequencies would be severely regularized. \revise{As a result, the first ML-EM iterations would mostly update the low frequencies, but  the high frequencies would also be slightly affected. 
As the algorithm proceeds, the structure would include more high frequency content, thus the variance of these frequency shells would increase, and the effective resolution of the 3-D map would gradually improve.} 
This strategy of initializing with a low resolution structure and gradually increasing the resolution is called ``frequency marching'' and is shared by many packages in the field; see further discussion in Section~\ref{sec:FM}. 
Another popular package, cryoSPARC, also employs an \revise{ML-EM} algorithm  with the prior that each voxel  in real space was drawn independently from an independent Poisson distribution with a constant parameter~\cite{punjani2017cryosparc}.  
\revise{We mention that such statistical priors (e.g., each voxel or frequency 
	is drawn from Poisson or normal distributions) are usually chosen out of mathematical and computational convenience   and may bias the reconstruction process. It is an open challenge to replace those statistical priors by priors which are based on or inspired by biological knowledge, such as previously reconstructed structures.}

The main drawback of the \revise{ML-EM} approach, \revise{especially at high resolution}, is the high computational load of the E-step in each iteration. Specifically, each \revise{ML-EM} iteration requires correlating each experimental image (typically a few hundred thousands) with multiple synthetic projections of the current structure estimate (sampled densely over $SO(3)$) and their 2-D translations.   
To alleviate the computational burden, a variety of methods are employed to narrow the 5-D search space. Popular techniques are based on \revise{multi-scale} approaches, where an initial search is done on a coarse grid followed by local searches on finer grids. 
A more sophisticated idea was suggested in~\cite{punjani2017cryosparc} based on  the ``branch and bound'' methodology that rules out  regions of the search space that are not likely to contain the optimum of the objective function.

%-------\emph{ab initio} modeling--------------------

\subsection{Ab initio modeling}
\label{sec:ab_initio_modeling}

In this section, we describe some of the intriguing ideas
that were proposed to construct \emph{ab initio} models, namely, models that do not require an initial guess.
These methods usually result in  low-resolution estimates that can be later refined as described in the previous section.
To emphasize the importance of robust \emph{ab initio} techniques, we begin by discussing \emph{model bias}---a phenomenon of crucial importance in cryo-EM and statistics in general.

%-------Model bias--------------------

\subsubsection{Model bias and validation}
\label{sec:model_bias}

The cryo-EM  inverse problem is non-convex and thus  
the output of the 3-D reconstruction algorithms may depend on their initializations. This raises the validation problem: how can we verify that a given estimate is a faithful representation of the underlying data?  
Currently, a 3-D model is treated as valid if its structural features meet the common biological knowledge (primary structure, secondary structure, etc.)  and if it passes some computational tests based on different heuristics. 
For instance, the reconstruction algorithm can be initialized from many different points. If the algorithm always converges to the same or similar structures, it strengthens the confidence in the attained solution. Moreover, since the data is  usually uploaded to public repositories, different researchers can examine it and compare the results against each other~\cite{rosenthal2015validating}; \revise{see a more detailed discussion in Section~\ref{sec:verification}.} 

Despite all precautions taken by researchers in the field, the verification methods are not immunized against systemic errors---this pitfall is dubbed {model bias}. One important example concerns the particle picking stage in which one aims to detect and extract  particle projections from  noisy micrographs. 
The majority  of the  detection algorithms in the field are based on  template matching techniques, despite their intrinsic flaws.  Specifically, choosing improper templates can lead to erroneous detection, which in turn biases the 3-D reconstruction algorithm toward the chosen templates; see~\cite{henderson2013avoiding} and further  discussion on particle picking techniques in~\ref{sec:particle_picking}.

The model bias phenomenon is exemplified by the ``Einstein from noise'' experiment~\cite{shatsky2009method}. 
In this experiment, $N$ images of pure i.i.d.\ Gaussian noise are  correlated with a reference image (in the original paper the authors chose an image of Einstein as the reference and thus the name). Then, each pure-noise image is  shifted to best align with the reference image (based on the peak of their cross-correlation), and finally all  images are averaged. 
Without bias, one would expect that averaging pure noise images will converge toward an image of zeros as $N$ diverges.
However, in practice the resulting image is similar to the reference image, that is, the algorithm is biased toward the reference image; see~\cite[Figure 2]{shatsky2009method}. 
In the context of cryo-EM, the lesson is that without prudent algorithmic design, the reconstructed molecular structure may reflect the scientist's presumptions rather than the structure that best explains the data.

We now turn our attention to some of the existing techniques for \emph{ab initio} modeling.

%-------Stochastic gradient descent--------------------

\subsubsection{Stochastic gradient descent over the likelihood function}

Stochastic gradient descent (SGD) gained popularity in recent years, especially due to its invaluable role in the field of deep learning~\cite{bottou2018optimization}. The underlying idea is very simple. Suppose that we wish to minimize an objective function of the form
\begin{equation}
f(x) = \frac{1}{N}\sum_{i=1}^N f_i(x). 
\end{equation}
A gradient descent algorithm is an optimization technique used to minimize the objective function by iteratively moving in the direction opposite to its gradient. 
Its $t^{\text{th}}$ iteration takes on the form
\begin{equation} \label{eq:GD}
x_{t+1} = x_{t} - \eta_t\frac{1}{N}\sum_{i=1}^N \nabla f_i(x_t), 
\end{equation}
where $\nabla$ denotes a gradient, and $\eta_t$ is called the step size, or learning rate.  
In SGD, we simply replace the full sum by a random element:
\begin{equation} \label{eq:SGD}
x_{t+1} = x_{t} - \eta_t \nabla f_i(x_t),
\end{equation}
where $i$ is chosen  uniformly at random from $\{1,\ldots,N\}$. In expectation, the direction of each SGD step is the direction of the full gradient~\eqref{eq:GD}.
The main advantage of SGD is that each iteration is significantly cheaper to compute than the full gradient~\eqref{eq:GD}. 
It is also easy to modify the algorithm to sum over few random elements at each iteration, rather than just one, to improve robustness at the cost of higher computational load per iteration.
In addition, it is commonly believed that the inherent randomness of  SGD  helps to escape local minima in some non-convex problems.  

This strategy was proven to be effective to construct \emph{ab initio} models using cryo-EM, achieving resolutions  of up to 10\AA~\cite{punjani2017cryosparc}. 
The objective function is the negative log of the posterior distribution. 
As the log posterior involves a sum over the experimental images~\eqref{eq:likelihood}, the SGD scheme chooses at each iteration one random image (or a subset of  images) to approximate the gradient direction.

%-------Common-lines--------------------

\subsubsection{Estimating viewing directions using common-lines}
\label{sec:common_lines}

If the orientations and locations of the particles in the ice are known (equivalently, the 3-D rotations and the 2-D translations in~\eqref{eq:model}), then recovering the 3-D structure is a linear problem that enjoys many solutions developed for CT imaging. 
In this section, we survey an analytical method to estimate the viewing directions using the common-lines property. 

Due to  the Fourier slice theorem, any pair of projection images has a pair of central lines (one in each image) on which their Fourier transforms agree. 
For generic molecular structures \revise{(with no symmetry)}, it is possible to uniquely identify this common-line, for
example, by cross-correlating all possible central lines in one image with all possible central lines in the other image, and choosing the pair of lines with maximum cross-correlation.

The common-line pins down two out of the three Euler angles associated with the relative rotation \revise{$R_i^{-1} R_j$} between
images $I_i$ and $I_j$.  A third image is required to determine the third angle: the three common-line pairs between three images uniquely determine their relative rotations. 
This technique is called \emph{angular reconstitution}, and it was suggested, independently, by Vainshtein and Goncharov~\cite{vainshtein1986determination} and Van Heel~\cite{van1987angular}. The main drawback of this procedure is its sensitivity to noise; it requires the three pairs of common lines to be accurately identified. 
Moreover, estimating the rotations of additional images sequentially (using their common-lines with the previously rotationally assigned images) can quickly accumulate errors. 
  
As a robust alternative, it was proposed to estimate the rotations $R_1,\ldots,R_N$ from the common-lines of all pairs \revise{$I_i,I_j$} simultaneously; this framework is called \emph{synchronization}. 
Since this strategy takes all pair-wise information into account, it has better tolerance to noise.
The rotation assignment can be done using a spectral algorithm or semidefinite programming~\cite{singer2011three} and enjoys some theoretical \revise{guarantees}; see for instance~\cite{singer2011angular}. 
Nevertheless, as discussed in Section~\ref{sec:computional_challenges}, in the low SNR regime any method for estimating the rotations would fail.
Therefore, the method requires  as input high SNR images that can be obtained using a procedure called \mbox{2-D} classification; see further discussion in Section~\ref{sec:2Dclassification}.
2-D classification, however, blurs the fine details of the  images and therefore the attained resolution of the method is limited~\cite{greenberg2017common}.

Three last comments are in order. First, if the structure possesses  a nontrivial symmetry, there are multiple common-lines between pairs of images that should be considered~\cite{pragier2019common}.   
Second, interestingly, this technique cannot work when the underlying object is a 2-D image (rather than 3-D as in cryo-EM). In this case, the Fourier transform of the tomographic projection (which is a 1-D function in that case) is a line that goes through the origin of the 2-D Fourier space of the underlying object. Therefore, the single common point of any two projections taken from different angles is the origin, and  thus there is no way to find the relative angle between them~\cite{basu2000uniqueness}.
\revise{Finally, the common-lines method is effective for additional signal processing applications, such as the  
	study of specimen populations~\cite{levis2018statistical}.}

%-------The method of moments (Kam's method)--------------------

\subsubsection{The method of moments} 

Suppose that a set of parameters $x$ characterizes a distribution $p(y|x)$ of a random variable~$y$ \revise{with $L$ entries}. We observe  $N$ i.i.d.\ samples of $y$ and wish to estimate $x$. 
In the method of moments, the underlying idea is to relate the moments of the observed data with $x$. Those moments can be estimated from the data by averaging over the observations: 
\begin{align} \label{eq:moments}
\frac{1}{N}\sum_{i=1}^Ny_i \approx\E{p\left(y| x\right)} &= M_1(x) \nonumber \\
\frac{1}{N}\sum_{i=1}^Ny_iy_i^T \approx  \E{p\left(yy^T|x\right)} &= M_2(x) \\
&\vdots \nonumber \\ 
\frac{1}{N}\sum_{i=1}^Ny_i^{\otimes k} \approx\E{p\left(y^{\otimes k}| x\right)} &= M_k(x),  \nonumber
\end{align}
\revise{where $y^{\otimes k}$ is a tensor  with $L^k$ entries and the entry indexed by 
$n=(n_1,\ldots,n_k)\in\mathbb{Z}_L^{k}$
is given by $\prod_{i=1}^ky(n_i)$.}
By the law of large numbers, when $N\to\infty$ the average converges almost surely to the expectation. 
The right-hand side underscores that the expectations are solely functions of $x$; the \revise{moment tensors $M_1,M_2,\ldots,M_k$} can be derived analytically.  
For instance, in cryo-EM, the expectation is taken against the random rotations, translations and the noise; the moments are functions of the 3-D volume, the 5-D distribution over the space of 3-D rotations and 2-D translations, and some parameters of the noise statistics (due to bias terms).
The final step of the method of moments is to solve the system of equations~\eqref{eq:moments}, which might be  highly non-trivial.
Importantly, since estimating the moments requires only one pass over the  observations, it can  be  done potentially on the fly  during data acquisition. 
This is especially important for cryo-EM in which hundreds of thousands of images need to be processed.

Computing the $q$th moment involves the product of $q$ noisy terms and thus 
the variance of its estimation scales as $\sigma^{2q}/N$.
As a result, when invoking the method of moments
it is crucial to determine the lowest-order moment that identifies the parameters $x$ uniquely:
this moment determines the  \emph{estimation rate} of the problem, that is, how many samples are required to accurately estimate the parameters. 
Remarkably, it was shown that under a general statistical model that is tightly-related to the cryo-EM problem (as will be described in Section~\ref{sec:mra}), the first moment that distinguishes between different parameters determines the \emph{sample complexity} of the problem (namely, how many observations are necessary to attain an accurate estimate) in the low SNR regime~\cite{abbe2018estimation}. 
In addition, the number of equations in ~\eqref{eq:moments} increases with the order of moments. Thus, using lower-order moments reduces the \emph{computational complexity}.

The first to suggest the method of moments for single particle reconstruction, 40 years ago, was Zvi Kam~\cite{kam1980reconstruction}.
Under the simplifying assumptions that the particles are centered, the rotations are uniformly distributed and by ignoring the CTF, he showed that the second moment of the experimental images determines the 3-D structure, up to a set of orthogonal matrices. 
To determine those matrices, he suggested to compute a subset of the third-order moment.  
Recently, it was shown that under the same assumptions the structure is indeed determined uniquely from the third moment~\cite{bandeira2017estimation}.  
Remarkably, if the distribution of rotations is non-uniform, then the second moment suffices~\cite{sharon2019method}.
Therefore, in light of~\cite{abbe2018estimation}, the sample complexity of the cryo-EM problem in the low SNR regime, under the specified conditions, scales as $\sigma^4$ and $\sigma^6$  for non-uniform and uniform distribution of rotations, respectively.
It is still an open question  how many images are required to solve the full cryo-EM problem; see further discussion in Section~\ref{sec:challenges_ahead}.

Interest in the method of moments has been recently revived, largely due to its potential application to X-ray
free electron lasers (XFEL)~\cite{liu2013three,von2018structure}. 
We believe that the method of moments has been \revise{largely} overlooked
by the cryo-EM community since Kam's paper \revise{for three} main reasons. 
\revise{First, Kam's original formulation required uniform distribution of viewing directions, an assumption that typically does not hold in practice; a recent paper extends the method to any distribution over $SO(3)$~\cite{sharon2019method}.}
\revise{Second}, estimating the second and third order statistics accurately requires a large amount of data that was not available until recent years. \revise{Third},  accurate estimation of high-order moments for high-dimensional problems requires modern statistical techniques, such as eigenvalue shrinkage in the spiked covariance model, that have only been introduced in the last decade~\cite{donoho2018optimal}.

%-------Frequency marching--------------------

\subsubsection{Frequency marching}
\label{sec:FM}

Most of the cryo-EM reconstruction algorithms start with a low-resolution estimate of the structure, which is then gradually refined to higher resolutions. This process is dubbed \emph{frequency marching}.
Frequency marching can be done explicitly by constructing a low-resolution model using an \emph{ab initio} technique that is later refined by \revise{ML-EM}~\cite{punjani2017cryosparc}, or more implicitly by an iteration-dependent regularization~\cite{scheres2012relion}. 

In~\cite{barnett2017rapid}, a deterministic and mathematically rigorous method to gradually increase the resolution was proposed. The 3-D Fourier transform of the object is expanded by concentric shells. At each iteration, each experimental image is compared with many simulated projections of the current low-resolution estimate, and a viewing direction is assigned to each experimental image by template matching (hard angular assignment). Given the angular assignments, an updated structure with one more frequency shell is computed by solving a linear least squares problem, thus progressively increasing the resolution. 

\section{Building blocks in the \revise{computational} pipeline}
\label{sec:building_blocks}

This section elaborates on some of the  building blocks in the algorithmic pipeline of single particle reconstruction using cryo-EM.
For each task, we introduce the problem and discuss the underlying principles behind some of the most commonly used solutions.

\subsection{Motion correction} \label{sec:motion_correction}

During data collection, the electron beam  induces sample motion   that \revise{mitigates} high-resolution information. 
The modern detectors (\revise{direct electron detectors}) acquire multiple frames per micrograph, allowing to partially correct the motion \revise{blur} by aligning and averaging the frames.
In essence, motion correction is an alignment problem (also referred to as registration or synchronization) that shares many similarities with classical tasks in signal processing and computer vision. The main challenge in \revise{alignment} is the high noise levels that hamper precise estimation of relative shifts between frames.

\begin{figure}
	\centering 
	\includegraphics[width=1\linewidth]{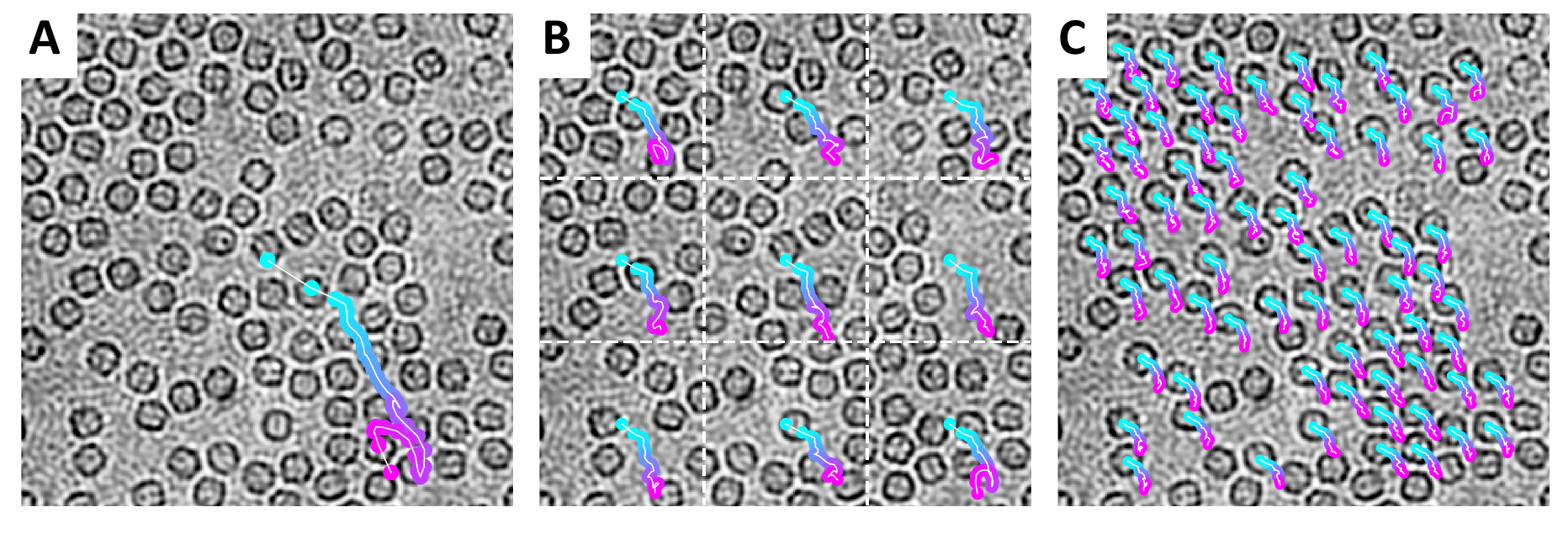}
	\caption{Beam-induced motion correction. All movement trajectories in this figure are shown color coded, using cyan to represent the position of first frame and magenta to represent the position of the last frame in the sequence. A) Strategies for global motion correction compensate for movement of the specimen across the entire field of view containing multiple particles~\cite{grant2015measuring}. B) Semi-local strategies for motion correction align frames across sub-regions defined on a discrete grid~\cite{zheng2017motioncor2}. C) Per-particle or local drift-correction allows accurate tracking of individual particles throughout the exposure to electrons~\cite{bartesaghi2018atomic}.}
	\label{fig:motion_correction}
\end{figure}

Several solutions were proposed to the motion correction problem based on a variety of methods, such as  pair-wise alignment of all frames and optical flow; see a survey on the subject in~\cite{ripstein2016processing} and references therein. The first generation solutions  aimed to estimate the movement of the entire micrograph, however,  the drift is not homogeneous across the entire field of view, motivating the development of more accurate local techniques.
One such strategy is implemented by the software MotionCor2~\cite{zheng2017motioncor2}, which describes the  motion as a block-based deformation that varies smoothly throughout the exposure.
The micrograph is first divided into patches, and motions within each patch are estimated based on cross-correlation. 
Then, the local motions are fitted to a time varying 2-D polynomial; 
the polynomial is quadratic in the 2-D frame coordinates and cubic in the time axis.
Finally, the frames are summed, with or without dose weighting which was determined  according to a radiation damage analysis. 
More recently, strategies for per-particle motion correction have been proposed that yield significant improvements in resolution~\cite{bartesaghi2018atomic,zivanov2019bayesian}.
Figure~\ref{fig:motion_correction} demonstrates
examples of different motion correction strategies.

\subsection{CTF estimation and correction} \label{sec:CTF}

Each cryo-EM image is affected by the PSF of the microscope through convolution with a kernel \revise{$h_i$}; see~\eqref{eq:model}.
The Fourier transform of the PSF---the CTF---suffers from multiple zero-crossings, making its inversion (i.e., deconvolution)  challenging. 
To compensate for the missing frequencies, the standard routine is to apply different PSFs to different micrographs. 
Thus, spectral information that was suppressed in one micrograph may appear in another. 

The CTF cannot be \revise{specified} precisely by the user. \revise{Instead,}
the number and location of the zero-crossings \revise{is given by the specific experimental} parameters  of the  microscope, which are hard to control. 
Thus, a preceding step entails estimating the CTF from the acquired data. 
Specifically, the CTF is modeled as \revise{(under the weak-phase approximation~\cite{wade1992brief})}
\begin{align} \label{eq:CTF_model}
\text{CTF}(k,\lambda,\Delta f, C_s) = -\sin \left(\pi \lambda |k|^2\Delta f - \frac{\pi \lambda^3 |k|^4C_s}{2}  + \alpha \right)\cdot E(|k|),
\end{align}
where $k$ is the frequency index, \revise{$\alpha$ is a small phase shift term,} $\lambda$ is the electron wavelength, $\Delta f$ is the objective defocus,  $C_s$ is the spherical aberration\revise{, and $E(|k|)$ is  an exponentially decaying envelope function---specified by a parameter called B-factor---due to the beam energy spread, the beam coherence, and the sample drift~\cite{rohou2015ctffind4,sorzano2007fast}}.
\revise{The CTF mitigates the very low frequencies, and therefore centering projection images (that is, finding their center of mass) is challenging. In fact,  the center of mass depends on the CTF, and thus two projection images with the same viewing angles but different CTFs (e.g., different defocus values) will have different centers in real space.
The CTF is approximately constant along concentric rings, although in practice those rings might be slightly deformed;} this deformation can be modeled by an additional parameter called astigmatism).
The \revise{defocus} value of the microscope has a pivotal role: high defocus values enhance low-resolution features and improve the contrast of the image (increase the SNR), whereas low defocus values enhance  high-resolution features (fine details) at the cost of lower contrast (lower SNR). 
 
CTF estimation begins by computing the power spectrum of the whole micrograph. The power spectrum exhibits ``rings'', called \emph{Thon rings}, that correspond to the oscillations of the sine function~\eqref{eq:CTF_model}; those rings are used to fit the CTF's parameters (see Figure~\ref{fig:1}B).
Two popular software packages CTFFIND4~\cite{rohou2015ctffind4} and Gctf~\cite{zhang2016gctf} estimate those parameters by generating multiple templates according to the postulated model~\eqref{eq:CTF_model}. 
The template that best fits the measured Thon rings is used as the estimated CTF. 

Once the CTF was estimated, the goal it to invert its action. 
The challenge stems from the structure of the CTF~\eqref{eq:CTF_model} \revise{because it has} many small values and thus direct inversion is impossible.   
The most popular technique for CTF correction, ``phase flipping'',  is embarrassingly simple: it disregards
the information about amplitude changes and corrects the data only for the sign of the CTF, and thus obtains the correct phases in Fourier space. 
Namely, if a micrograph is represented in Fourier domain by 
\begin{align}
I_{\text{micrograph}}(k) = \text{CTF}(k) I(k),
\end{align}
where $I(k)$ is the micrograph before the CTF  action,
then the corrected image would be 
\begin{align}
I_{\text{corrected}}(k) = \text{sign}(\text{CTF}(k)) I_{\text{micrograph}}(k) = \vert \text{CTF}(k)\vert  I(k).
\end{align}
Another approach is to apply Wiener filtering that attempts to \revise{correct} both the amplitudes and the \revise{phases~\cite{frank1995correction,sindelar2011adaptation}}. 
However, this approach requires the knowledge of the SNR (which is not trivial to estimate in a low SNR environment) and cannot correct for missing frequencies. 

Instead of inverting the CTF explicitly, an attractive alternative is to incorporate the CTF into the forward model of the reconstruction algorithm. This can be done naturally in the \revise{ML-EM} approach, the method of moments, and the SGD framework. 
Because different CTFs are applied to different micrographs, in principle,  full CTF correction is viable.

\subsection{Particle picking} \label{sec:particle_picking}

A micrograph consists of regions which contain only noise, regions with noisy 2-D
projections, and  regions with other contaminants, such as carbon film. 
In particle picking, the goal is to detect and extract the 2-D projections (particles) from the noisy micrographs.  
High-resolution reconstruction typically requires hundreds of thousands of particles, and thus manual picking is time consuming and tedious. In addition, it may introduce \revise{subjective} bias into the procedure.

Many solutions were proposed in the literature for the particle picking problem, based on standard edge detection techniques, machine learning~\cite{wang2016deeppicker,zhu2017deep,wagner2019sphire}, and template matching. For the latter, a  popular strategy is the one implemented in RELION. In this framework, the
user manually selects a few hundreds of particles. These particle images
are then 2-D classified (see next section), and the 2-D classes are used as templates for an automatic particle picking based on 
template matching~\cite{scheres2015semi}. 

As discussed in Section~\ref{sec:model_bias}, a major concern of any particle picking algorithm is a systematic  bias. 
To alleviate the risk for model bias, a fully automated technique was proposed in~\cite{heimowitz2018apple}.
Rather than templates, the algorithm selects automatically a set of reference windows that include both particle and noise windows. The selection is based on the local mean and variance: regions with particles typically have lower mean and higher variance than regions without particles.  Then, regions in the micrograph are correlated with the reference windows. The regions which are most likely to contain a particle (high correlation) and those which are least likely (low correlation with all reference windows) are used to train a support vector machine classifier. The output of this linear classifier is used to pick the particles.

\revise{In practice, a considerable number of picked particles are usually non-informative, containing  adjacent particles  that are too close to each other, only part of a particle, or just pure noise.
Consequently, it is common to try and prune out these outliers. In~\cite{sanchez2018deep}, it was proposed to employ several particle pickers simultaneously and compute a consensus between them using a deep neural network. In~\cite{zhou2019unsupervised}, a simple pruning is devised by viewing the output of the particle picking algorithm as a mixture of Gaussians.}

\subsection{2-D classification} \label{sec:2Dclassification}

As experimental images corresponding to similar viewing directions  tend to be very much alike (due to the smoothness  of the molecular structure), 
it is common to divide  the images into several classes (i.e., clustering) and average them  to increase the SNR. 
This process is called 2-D classification and 
the averaged images are referred to as ``class averages.''
Since the global in-plane rotation of each micrograph is  arbitrary, the clustering should be invariant under in-plane rotations: two images from a similar viewing direction but with a different in-plane rotation are supposed to be grouped together after appropriate alignment.
The class averages are used for a variety of tasks: to construct \emph{ab initio} models, as templates for particle picking, to provide a quick
assessment of the particles, to remove picked particles which are associated with non-informative classes, and for symmetry detection. 

There are three main computational aspects that make the 2-D classification task quite challenging. First, as already mentioned, low SNR impedes accurate clustering and alignment. Second, the high computational complexity of finding the optimal clustering among hundreds of thousands of images may be prohibitive unless designed carefully. Third, it is not clear what is the appropriate similarity metric to accurately compare images. 

\revise{Many different strategies for 2-D classification were proposed; see for instance~\cite{van1981use,sorzano2010clustering,reboul2016stochastic,huang2016robust}. }
A popular 2-D classification algorithm, implemented in RELION~\cite{scheres2005maximum}, is based on the \revise{ML-EM} scheme.
In this algorithm, each observed image is modeled as a sample from the statistical model
\begin{align}
y_i &= h_i\ast T_{t_i}R_{\theta_i} x_{k_i}+\varepsilon_i, & i=1,\ldots,N,
\end{align}
where $k$ is distributed uniformly over $\{1,\ldots,K\}$ (these are the $K$ class averages to be estimated), $\theta$ is distributed uniformly over $\theta\in[0,2\pi)$, $t$ is drawn from an isotropic 2-D Gaussian, \revise{$h_i$} is the estimated PSF, and $\varepsilon_i\sim\mathcal{N}(0,\sigma^2I)$.  Given this model, it is straight-forward to implement an \revise{ML-EM} algorithm to estimate $x_1,\ldots,x_K$ following the guidelines of Section~\ref{sec:3D_reconstruction}.
This algorithm, albeit popular, suffers from three imperative weaknesses.  First, the computational burden of running \revise{ML-EM} is a major hurdle  as the algorithm needs to go through all experimental images at each iteration. Second,  it assumes that each experimental image is originated from only $K$ possible class averages. However, this is an inaccurate model as the orientation of  particles in the ice layer are distributed continuously over $SO(3)$ \revise{(in addition to structural variability that may be present)}. 
Finally, \revise{ML-EM} typically suffers from the ``winner takes  all'' phenomenon: most experimental images would correlate well with, and thus be assigned to  \revise{the} class averages that enjoy higher SNR. As a result, \revise{ML-EM} tends to output only a few, low-resolution classes. 
\revise{This phenomenon was already recognized by~\cite{sorzano2010clustering} and is also present for  ML-EM based 3-D classification and refinement.} 

An alternative solution was proposed in~\cite{zhao2014rotationally}. In this framework, each image is averaged with its few nearest neighbors, after a proper in-plane alignment. The nearest neighbor search is executed efficiently over the bispectra of the images. The bispectrum is a rotationally-invariant feature of an image--- that is, it remains unchanged under an in-plane rotation.
To reduce the computational complexity and  denoise the data, each image is first compressed using a dimensionality reduction technique---called steerable PCA---which is the main topic of the next section.

\subsection{Denoising and dimensionally reduction techniques } \label{sec:PCA}

Principal component analysis (PCA) is a widely used tool for linear dimensionality reduction and denoising, \revise{dating back to  Pearson~\cite{pearson1901liii,jolliffe2011principal}}. 
PCA computes the best (in the least squares sense) low-dimensional  affine space that approximates the data by projecting the images into the space spanned by the leading eigenvectors of their covariance matrix.
\revise{In cryo-EM, covariance estimation  was introduced by Kam~\cite{kam1980reconstruction} and used for dimensionality reduction and image classification  for the first time by  van Heel and Frank~\cite{van1981use}. PCA is} often used to denoise the experimental images and as part of the 2-D classification stage. 

Cryo-EM images are equally likely to appear in any in-plane rotation.  
Consequently, when performing PCA, it makes sense to take all in-plane rotations of each image into account. Luckily, there is a simple way to account for all in-plane rotations without rotating the images explicitly. This can be done by expanding the images in  a \emph{steerable basis}: a basis formed as an outer product of radial functions with \revise{an angular}  Fourier basis; examples are Fourier Bessel, 2-D prolate spheroidal wave functions, and Zernike polynomials. 
When integrating over all possible in-plane rotations, the  covariance matrix  of the expansion coefficients enjoys  a block-diagonal structure, reducing the computational load significantly: while the covariance matrix of images of size $L\times L$ has~$L^4$ entries, the  block diagonal  structure guarantees  that merely
 $O(L^3)$ entries are non-zero. 
This in turn reduces  the computational complexity of  \emph{steerable PCA} from $O(NL^4 + L^6)$ to $O(NL^3 + L^4)$: the first term is the cost of computing the sample covariance over $N$ images, and the second term is the cost of the eigendecomposition over all blocks~\cite{zhao2016fast}. 
%The methodology of steerable PCA was recently extended to non-linear  dimensionality reduction, by harnessing the fact that all projection images lies upon a 3-D manifold (diffeomorphic to the group $SO(3)$)~\cite{landa2018steerable}. 

Classical covariance estimation techniques usually assume that the number of samples is considerably larger than the signal's dimension. However, this is not the typical case in cryo-EM in which  the  dimensionality of the molecule is of the same order as the number of measurements. 
In this regime, one can take advantage of recent developments in high-dimensional statistics under the ``spiked covariance model''~\cite{donoho2018optimal}. 
These techniques, based on eigenvalue shrinkage, were recently applied successfully to denoise cryo-EM images~\cite{bhamre2016denoising}. 

\section{Mathematical frameworks for cryo-EM data analysis}
\label{sec:abstract_models}

Inspired by the cryo-EM problem, a couple of abstract \revise{mathematical} models were studied in recent years. These models provide a general \revise{framework for analyzing} cryo-EM from  theoretical and statistical perspectives,  while removing some of its complications. In what follows, we introduce the models and succinctly describe some intriguing results.

\subsection{Multi-reference alignment}
\label{sec:mra}

\begin{figure}
	\centering 
	\includegraphics[width=.9\linewidth]{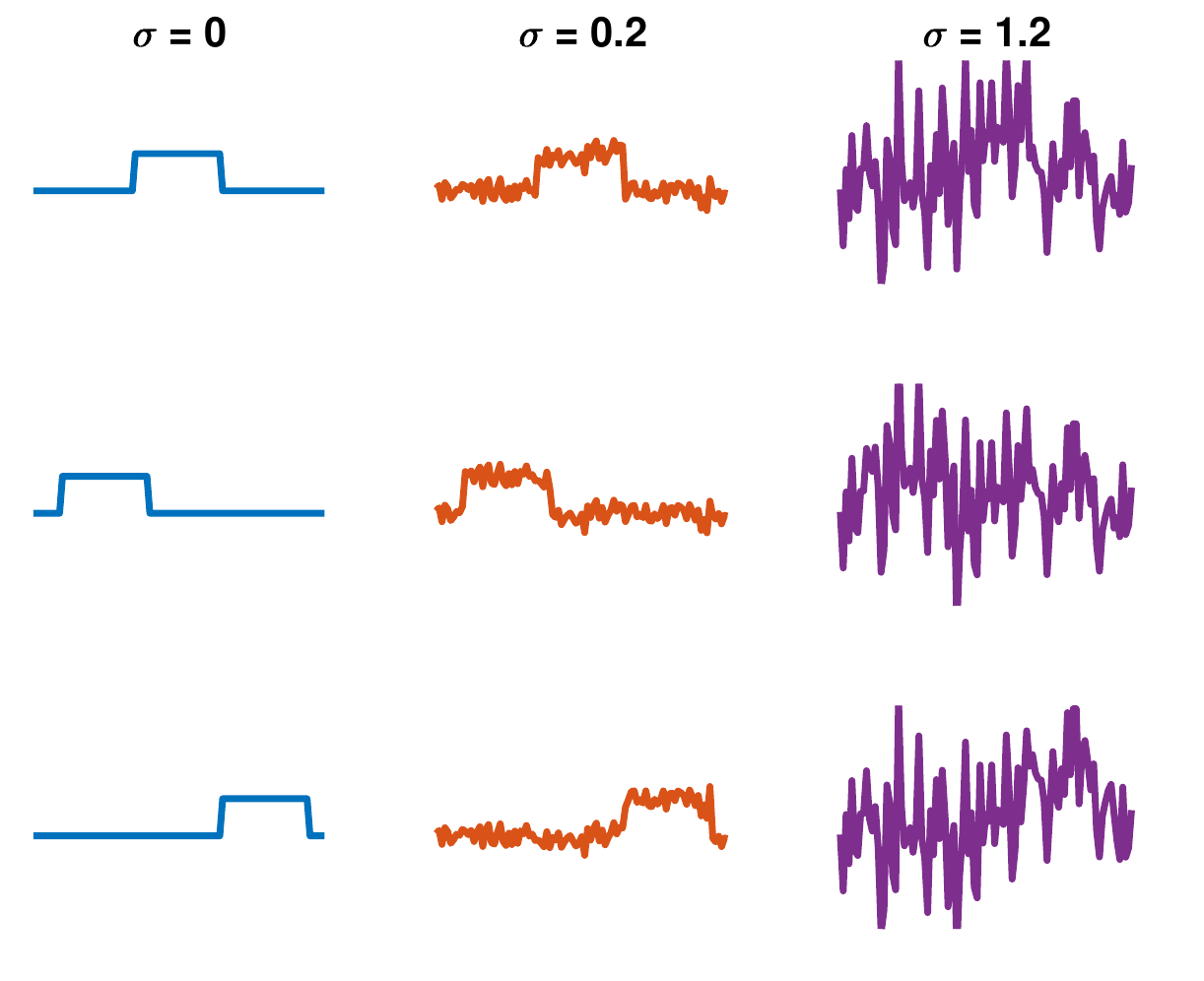}
	\caption{An example of multi-reference alignment observations at different noise levels $\sigma$. Each column consists of three different cyclic shifts (the group actions) of a 1-D periodic discrete signal (the linear operator $T$ is the identity operator). Clearly, if the noise level is low, estimating the signal is easy: one can align the observations (i.e., undo the group action) and average out the noise. The  challenge  is to estimate the signal when high noise levels hinder alignment (right column). 
	}
	\label{fig:MRA}
\end{figure}

The multi-reference alignment (MRA) model reads~\cite{bandeira2015non}:
\begin{align} \label{eq:mra}
y_i &= T_i(g_i\circ x) + \varepsilon_i,  & g_i\in G, 
\end{align}
where $G$ and \revise{$T_i$} are, respectively, a known compact group and  linear \revise{operators}, and ${\varepsilon_i\sim\mathcal{N}(0,\sigma^2I).}$ The signal $x\in\R^L$ is assumed to lie in a known space (say, the space of signals with a finite spectral expansion) that random elements of $G$ act upon.
\revise{The  task is  to estimate the signal $x$ from the observations $y_1,\ldots,y_N$, while the group elements $g_1,\ldots,g_N$ are unknown. The recovery is possible
  up to left
multiplication by some group element $g\in G$---namely, we wish to estimate the orbit of $x$ under the group action.}
Figure~\ref{fig:MRA} shows an example of three discrete 1-D MRA observations under the group of cyclic shifts $\mathbb{Z}/L$ at different noise levels.
If we assume perfect particle picking, then the cryo-EM model~\eqref{eq:model} is a special case of~\eqref{eq:mra} when $G$ is the group of 3-D rotations $SO(3)$ and $T$ is the linear operator that takes the rotated structure, integrates along the z-axis (tomographic projection), convolves with the PSF, and samples it on a Cartesian grid. 
\revise{The MRA model formulates  many additional applications, including 	structure from motion in computer vision~\cite{agarwal2011building},  localization and mapping in robotics~\cite{rosen2019se},
	study of specimen populations~\cite{levis2018statistical}, optical and acoustical trapping~\cite{elbau2019inverse}, and denoising of permuted data~\cite{pananjady2017denoising}.
}

While surveying all recent results about the MRA model is beyond the scope of this article, we wish to present a remarkable information-theoretic result about the sample complexity of the problem (and therefore also for cryo-EM under the stated assumptions). 
 \revise{ In many cases, among them when $T=I$ and the cryo-EM setup (without shifts so that $G=SO(3)$),  one can  estimate the group elements $g_i$ from the measurements $y_i$ in the high SNR regime} using one of many synchronization techniques (based on spectral methods, semidefinite programming, or non-convex programming)~\cite{singer2011angular,singer2011three}. Once the group elements were identified,  
one can estimate the signal by aligning all  observations (undo the group actions) and averaging.
The variance of averaging over i.i.d.\ Gaussian variables is proportional to $1/(N\cdot\text{SNR})$---namely, the number of measurements should be proportional to $1/$SNR to achieve an accurate estimate. In this case, we say that the sample complexity of the problem in the high SNR regime is $1/$SNR: this is a common scenario in many data processing tasks. 
In the low SNR regime the situation is radically different.
Specifically, it was shown that when $\sigma\to\infty$, the sample complexity of the MRA problem is determined by the first moment $\bar{q}$ that distinguishes between different signals. 
More precisely, in the asymptotic regime where N and $\sigma$ diverge, the estimation error of any method is bounded away from zero if $N\cdot\text{SNR}^{\bar{q}}$ is bounded
from above. We then say that the sample complexity in the low SNR regime is proportional to $1/\text{SNR}^{\bar{q}}$~\cite{abbe2018estimation}. 

For the cryo-EM setup, assuming perfect particle picking and no CTF, it was shown that the sample complexity depends on the distribution of rotations: for uniform distribution the sample complexity scales as  $1/\text{SNR}^3$ (that is, the structure is determined by the third moment), while for generic non-uniform distribution the second moment suffices, and thus the sample complexity scales as $1/\text{SNR}^2$~\cite{bandeira2017estimation,sharon2019method}. Similar results were derived for simpler setups, for instance, when a discrete signal is acted upon by cyclic shifts (as shown in Figure~\ref{fig:MRA})~\cite{bendory2018bispectrum,abbe2018multireference,bandeira2017estimation}. 

The pivotal role of moments in sample complexity analysis led naturally to the design of algorithms based on the method of moments. 
Some of these algorithms hinge on  the tightly-related notion of \emph{invariant polynomials}.
A polynomial $p$ is called invariant under the action of a group $G$ if for any signal $x$ in the specified space and any $g\in G$, it satisfies $p(g\circ x) = p(x)$.
One particular important invariant is the third-order polynomial, the \emph{bispectrum}---originally proposed by John Tukey~\cite{tukey1953spectral}---that was used for 2-D classification~\cite{zhao2014rotationally} and \emph{ab initio} modeling~\cite{kam1980reconstruction}.
We refer the readers to~\cite{singer2018mathematics} and references therein for a more detailed account on the MRA problem.

\subsection{Multi-target detection}
\label{sec:mtd} 
 
Multi-target detection (MTD) is the problem of estimating a signal that occurs multiple times at unknown locations in a noisy measurement. 
In its simplest form, the MTD problem is an instance of the \revise{1-D} 
\emph{blind deconvolution} problem and can be written as
\begin{align} \label{eq:mtd}
y = x \ast s + \varepsilon, 
\end{align}
where $x\in\R^L$ is the target signal,  $s\in\{0,1\}^{N-L+1}$ is a binary signal whose 1's indicate the location of the signal copies in the measurement $y\in \R^N$, and $\varepsilon\sim \N(0,\sigma^2I)$. 
\revise{Detecting the signal occurrences in the data (that is, estimating the signal $s$) is the analog of particle picking in  cryo-EM.
Clearly, if the noise level is low, one can estimate $s$ (analogously, detect the particle projections in the micrograph), extract the signal occurrences, and average them to suppress the noise.} 
However,  as aforementioned, low SNR precludes detection \revise{(and particle picking)} and therefore one must estimate $x$ directly, without explicit estimation of $s$. 
In~\cite{bendory2019multi}, it was shown that under  certain generative models of $s$,  the signal $x$ can be  estimated provably, at any noise level, from the bispectrum (third-order statistics).
More ambitiously, numerical experiments suggest that the bispectrum suffices to estimate multiple signals $x_1,\ldots,x_K$ from a mix of blind deconvolution problems
\begin{align} \label{eq:mtd_heter}
y &= \sum_{i=1}^Kx_i \ast s_i + \varepsilon,
\end{align}
without explicit estimation of the binary signals $s_1,\ldots,s_K$ that indicate the location of the corresponding signals.
	
\begin{figure}
	\centering 
	\includegraphics[width=1\linewidth]{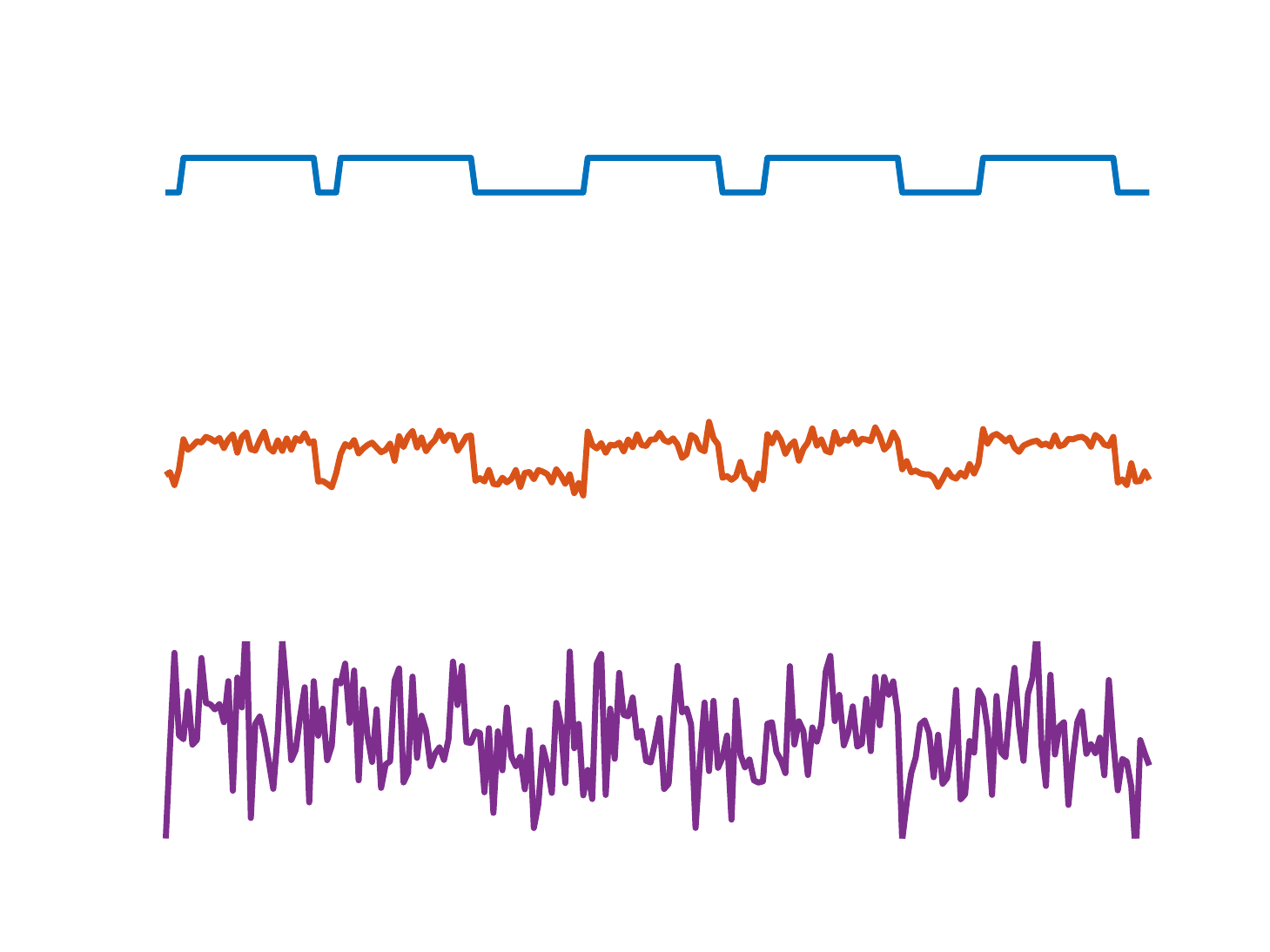}
	\caption{An example of a multi-target detection observation with five signal occurrences at different noise levels~\eqref{eq:mtd}. From top to bottom: $\sigma=0$, $\sigma=0.2$, and $\sigma=1.2$. When the noise level is low, it is easy to detect the signal occurrences and  estimate the underlying signal by averaging out the noise. 
	However, when the noise level is high (bottom), reliable detection is rendered challenging. The task is then to estimate the underlying signal directly,  without intermediate detection of its occurrences.
	}
	\label{fig:MTD}
\end{figure}

The MTD model can be  extended to formulate a  generative model of a micrograph:
\revise{the key is to treat $x$ as a random signal that represents the tomographic projections of the molecule $\phi$, rather than a deterministic signal as in~\eqref{eq:mtd} and~\eqref{eq:mtd_heter}.  
Specifically, locations  are chosen in the 2-D plane: these are  the positions of the embedded samples in the micrograph. 
For each location, a signal is drawn from a probability distribution described by the model
\begin{equation}
x = h\ast PR_\omega \phi,
\end{equation}
where the 3-D rotation $R_\omega$ is applied to the volume $\phi$ according to
a (possibly unknown) distribution of $\omega$ over $SO(3)$,  
$P$ is a tomographic projection, and $h$ is the microscope's PSF; the goal is to estimate $\phi$.
Therefore, the MTD model paves the way to fully model the cryo-EM problem, including most of its important features~\cite{bendory2019multi}. 
In particular, MTD provides a mathematical and computational framework for reconstructing $\phi$ directly from micrographs, without intermediate particle picking~\cite{bendory2018toward}. 
A full analysis of this model is still lacking.}

\section{Remaining computational and theoretical challenges}
\label{sec:challenges_ahead}

To conclude the article, we would like to discuss some of the interesting  and important computational and theoretical challenges  ahead for single particle reconstruction using cryo-EM.

\subsection{Conformational heterogeneity: modeling and recovery}
\label{sec:heterogeneity}

One of the important opportunities offered by cryo-EM is its ability to analyze different functional and conformational states of macromolecules. 
Mathematically, it entails  estimating multiple 3-D structures simultaneously; we refer to this problem as the  \emph{heterogeneity problem}.
There is no consensus about the proper way to model the heterogeneity problem and the  computational tools are not well-established.
We therefore believe that it offers  an opportunity for researchers with strong mathematical and computational background to make a profound impact on the field of structural biology.

We now briefly survey different approaches to handle the heterogeneity problem.
To this end, we extend the basic cryo-EM model~\eqref{eq:model} to account for conformational variability: 
\begin{align} \label{eq:model_heter_cont}
y_i &= h_i\ast T_{t_i} P R_{\omega_i}\phi_{i} +  \text{``noise''}, & i=1,\ldots,N,
\end{align} 
where the goal is to estimate the distribution from which the structures $\phi_{i},\ldots,\phi_{N}$ are sampled from. 
\revise{Using this formulation, the problem is ill-posed: there is not enough information in $N$ 2-D images to recover $N$ 3-D structures. Hence, additional assumptions on the structures must be made.}
The different techniques in the field can be  broadly classified into two categories: discrete and continuous models. 
The \emph{discrete heterogeneous} model assumes that the measurements stem from a few independent volumes (that is, a discrete distribution). Under this model, each projection image can be written as 
\begin{align} \label{eq:model_heter}
y_i &= h_i\ast T_{t_i} P R_{\omega_i}\phi_{k_i} +  \text{``noise''}, & i=1,\ldots,N,
\end{align} 
where $\phi_{k}, \, k=1,\ldots,K,$ represent the $K$ different volumes. The advantage of this approach is \revise{evident}: it is easy to extend the \revise{ML-EM} framework to incorporate several volumes simultaneously. Alternatively, some software packages apply a preliminary 3-D  classification stage, in which the 2-D experimental images are being classified into  different structures. CryoSPARC runs multiple trials of an SGD algorithm from different initializations. This procedure occasionally leads to multiple  low-resolution estimates, and each of those is refined by an \revise{ML-EM} algorithm~\cite{punjani2017cryosparc}. 
While the simplicity of the discrete model is a big advantage,  its drawbacks are apparent: it does not scale well for large $K$ and ignores the correlation between different functional states of the molecule and thus overlooks important information.

The second approach, referred to as \emph{continuous heterogeneity}, assumes that $\phi_1,\ldots,\phi_{N}$ can be embedded in a low-dimensional space.
For instance, one approach is to assume that the set of conformations lies in a linear subspace (that can be learned using PCA),  or in  more intricate low-dimensional manifolds (that could be learned by other spectral methods, such as diffusion maps).
An alternative suggestion was to model the structure as a set of rigid domains that can move one with respect to the other.
We refer the reader to a thorough survey  on the subject in~\cite{sorzano2019survey}.

\subsection{Verification} \label{sec:verification}

Given a 3-D reconstruction, how do we verify that it is a reliable and faithful representation of the underlying molecule?  This is a question of crucial importance for any scientific field. \revise{Nevertheless, there are no agreed upon  computational verification methods for cryo-EM.}
\revise{Several validation techniques were proposed in the cryo-EM literature. For example, it is possible 
to check the consistency of the 3-D map by recording pairs of images of the same  particles at different tilt angles and comparing the relative angle between orientations assigned to each projection: ideally it should agree with the  relative rotation angle of the microscope's specimen holder used during the experiment. This approach is called  the tilt-pair validation technique and is useful only for intermediate resolution structures~\cite{henderson2011tilt,russo2014robust,rosenthal2015validating}. In practice, structure validation} is based on a set of heuristics and the experts'  knowledge and experience. 
For instance, it is  common to initialize  3-D reconstruction algorithms from multiple random points; if all instances attain similar structures, it servers as a validation fidelity. 
If the same molecule was reconstituted by other technologies, such as X-ray crystallography and NMR, then  structures can be compared. 
In addition, experimental data is usually uploaded to public repositories and thus other researchers can process the same data, experiment with different computational techniques,  and compare the results~\cite{rosenthal2015validating}. 

Verification is also related to the question of determining the resolution of a recovered structure. The current convention is to reconstruct two structures independently, each from one half of the data. \revise{The two subsets are chosen at random.} 
The highest frequency for which the two structures agree (up to some tolerance) determines the resolution~\cite{scheres2012prevention,sorzano2017review}. 
This process can be understood as an indication for the confidence we have in the structure at a given resolution. However, this method is susceptible to systematic flaws; if the same \revise{refinement} procedure is applied to both halves of the data, it can induce correlated blunders (although the data is independent) and the resolution determination  would be unsound.
\revise{In~\cite{chen2013high}, 
a simple computational procedure was proposed to validate a structure by assessing the amount of overfitting that is present in the  3-D map.}
Establishing computational tools that provide confidence intervals to estimated structures and are immunized against systematic errors  is one of the remaining challenges in the field.

\subsection{Theoretical foundations of cryo-EM}

Many mathematical and statistical  properties  
 of the cryo-EM problem  
are still unexplained (and even unexplored). 
A prominent example is the sample complexity of the  full cryo-EM problem: 
given a fixed SNR level (that might be very low) and a fixed setup,  how many particles are required to achieve a desired resolution? An initial analysis was conducted in~\cite{bandeira2017estimation,sharon2019method}; see Section~\ref{sec:mra}.

Another interesting question concerns computational complexity and information-computational gaps~\cite{bandeira2018notes}. The fact that there is enough information to solve a problem (sample complexity) does not immediately imply that there is an efficient (polynomial time) algorithm to solve it. 
For example, it might be that for non-uniform distribution of rotations the second moment would suffice to estimate a 3-D structure from an information-theoretic perspective~\cite{sharon2019method}, but at the same time it could be computationally hard in that regime. In that case, we might  need to consider the third-order moment in order to design a computationally efficient algorithm.
Information-computational gaps have been observed empirically in  variants of the MRA and MTD models~\cite{boumal2018heterogeneous,bendory2019multi}.
Related questions---that might be even more challenging---deal with the properties of specific algorithms, such as \revise{ML-EM}, the method of moments, and  SGD  that are not well-understood (see for instance~\cite{balakrishnan2017statistical}).

Another interesting research thread regards the size limit of molecular structures that can be elucidated by cryo-EM. 
The common belief in the  community is that very small molecules  cannot be visualized using cryo-EM. The logic is simple: small molecules induce low contrast, and thus low SNR on the micrograph, which in turn hinders detection (particle picking)~\cite{henderson1995potential}. A recent paper contends this belief and suggests that it is possible, at least in principle, to reconstruct structures directly from the micrograph, without particle picking and at any SNR level, given enough data~\cite{bendory2018toward}.

\subsection{Machine learning}

The groundbreaking advances in machine learning in the last decade reshaped dramatically many computational fields and penetrated into some scientific applications. 
Naturally, learning techniques based on deep neural networks have been applied to cryo-EM as well.
\revise{Examples} are particle picking~\cite{wang2016deeppicker,zhu2017deep,wagner2019sphire} (as discussed in Section~\ref{sec:particle_picking}), \revise{validation~\cite{avramov2019deep}, 3-D reconstruction~\cite{zhong2019reconstructing}, and 
 particle pruning~\cite{sanchez2018deep}.}  
\revise{In addition, manifold learning techniques were designed for 3-D heterogeneity analysis~\cite{frank2016continuous,moscovich2019cryo} and deniosing~\cite{landa2018steerable}.}
Modern learning techniques were also implemented to other cryo-EM applications that do not involve single particle reconstruction; see for instance an application to feature extraction in cellular electron cryotomography~\cite{chen2017convolutional}.

Deep learning gained its popularity in applications with low noise levels. 
The performance of these techniques in more challenging environments, such as highly contaminated data, is not clear yet. 
In addition, supervised learning techniques are susceptible to model bias---the reconstruction will depend heavily on the training data, rather than on the experimental images.
This explains why the impact of deep learning on the cryo-EM field---especially on the more involved tasks, such as 3-D reconstruction and the heterogeneity problem---is limited at the moment.
We expect that more efforts in this direction will be made in the coming years.
In particular, it is still to be clarified whether this set of computational tools can outperform  current tools in the field, which are based on more classical statistics.

\section{Perspective} \label{sec:perspective}

Single particle reconstruction using cryo-EM is an alluring research area for researchers
interested in developing modern computational tools and sophisticated mathematical models to an
emerging scientific field.
In this article, we have introduced the problem of
constituting  3-D molecular structures using cryo-EM and described its unique computational characteristics and challenges.
We have delineated  relations between the cryo-EM inverse problem and  
a variety of disciplines at the core of signal processing, information theory, statistics, machine learning, and group theory. 
We believe that contributions from these areas have the potential to drive the field forward.
New ideas and solutions can, and should, be tested on experimental cryo-EM datasets publicly available online~\cite{empiar,emdb}.
We have also \revise{reviewed} two abstract frameworks to conveniently  study the cryo-EM inverse problem from computational, statistical and theoretical perspectives.  

We believe that the challenges arising in cryo-EM research provide an ample of opportunities to investigate and test novel algorithms and advanced mathematical techniques in order to impact a task of paramount importance: to broaden our understanding of fundamental mechanisms of life.

\section*{Acknowledgment}
\revise{ We thank the anonymous reviewers and the editor for their valuable comments and suggestions.}
AS was supported in part by Award Number R01GM090200 from the
NIGMS, FA9550-17-1-0291 from AFOSR, Simons Foundation Math+X Investigator
Award, the Moore Foundation
Data-Driven Discovery Investigator Award, and NSF BIGDATA Award IIS-
1837992.

\bibliographystyle{ieeetrans}

\end{document}